\documentclass{caps}
\usepackage{psfig}
\def\mk{}

\begin{document}

\def\bu{\noindent $\bullet$~}

\pagenumbering{roman}
%\tableofcontents
%\cleardoublepage
%\clearpage

\pagenumbering{arabic}

\author[Dimitrios Psaltis]
{DIMITRIOS PSALTIS\\Department of Physics, University of Arizona,
   Tucson, AZ 85721 USA}

\chapter{Accreting neutron stars and black holes:\newline
 a decade of discoveries}

\section{Introduction}

Since their discovery in 1962 (Giacconi et al.\ 1962), accreting
compact objects in the galaxy have offered unique insights into the
astrophysics of the end stages of stellar evolution and the physics of
matter at extreme physical conditions. During the first three decades
of exploration, new phenomena were discovered and understood, such as
the periodic pulsations in the X-ray lightcurve of spinning neutron
stars (Giacconi et al.\ 1971) and the thermonuclear flashes on 
neutron-star surfaces that are detected as powerful X-ray bursts
(see, e.g., Grindlay et al.\ 1976; \S3)\mk. Moreover, the masses of
the compact objects were measured in a number of systems, providing the
strongest evidence for the existence of black holes in the universe
(McClintock \& Remillard 1986; \S4)\mk.

During the last ten years, the launch of X-ray telescopes with
unprecedented capabilities, such as {\em RXTE}, {\em BeppoSAX}, the
{\em Chandra\/} X-ray Observatory, and {\em XMM-Newton} opened
new windows onto \mk the properties of accreting compact objects.
Examples include the rapid variability phenomena that occur at the
dynamical timescales just outside the neutron-star surfaces and the
black-hole horizons (van der Klis et al.\ 1996; Strohmayer et al.\
1996; \S2, \S4\mk) as well as atomic lines that have been red- and blue-shifted by
general relativistic effects in the vicinities of compact objects
(Miller et al.\ 2002b; Cottam et al.\ 2001). Accreting neutrons stars
and black holes have been monitored in broad spectral bands, from the
radio to $\gamma$-rays, leading to the discovery of highly
relativistic jets (Mirabel \& Rodriguez 1994; \S9\mk), to the indirect imaging
of the accretion flows (Horne 1985; \S5\mk), and to the possible
identification of neutron stars with masses close to the maximum value
allowed by general relativity (Barziv et al.\ 2000). Finally, the
theoretical modeling of accretion flows also experienced significant
advances, such as the identification of a whole suit of stable
solution for accretion flows beyond the standard model of
geometrically thin accretion disks (e.g., Narayan \& Yi 1994) and of
the most promising avenue towards explaining the very efficient
transport of angular momentum in accretion flows (e.g., Balbus \&
Hawley 1991).

The aim of this chapter is to provide a general overview of these
recent advances in the astrophysics of X-ray binaries in our galaxy.
The basic concepts have been reviewed in a number of textbooks (e.g.,
Shapiro \& Teukolsky 1983; Glendenning 2003) and review articles
(e.g., White, Nagase, \& Parmar 1995) and will only be briefly
mentioned here\mk. Several other classes of compact stellar X-ray
sources that do not involve accretion onto a neutron star or black
hole will also not be discussed in this chapter but are reviewed
elsewhere in this volume. These systems include: Isolated neutron
stars (\S8); Cataclysmic variables (CVs; \S10); Supersoft sources
(SSS; \S11); Soft Gamma-ray Repeaters and Anomalous X-ray Pulsars
(SGRs and AXPs; \S16); and Gamma ray bursts (GRBs; \S15). Finally,
accreting compact objects in other galaxies will be reviewed in \S11.

\subsection{X-ray binary systems}

Whether a compact object in a binary system is accreting mass in a
stable long-lived phase or not depends mostly on the mode of mass
transfer, the ratio of the mass of the compact object to that of the
companion star, and their orbital separation. For example, in the case
of a neutron star (with a mass $\sim 1.4-2.0$~$M_\odot$), stable mass
transfer through the inner Lagrangian point occurs only when the
companion fills its Roche lobe and has a mass smaller than that of the
neutron star. In such systems, mass is driven by angular momentum
loses due to gravitational radiation (for very small masses and
orbital separations) and magnetic breaking (for orbital periods $\le
2$~day) or by the evolution of the companion star (for orbital periods
$\ge 2$~day). These sources are significantly brighter in the X-rays
than in the optical wavelengths, with the flux at the latter spectral
band being mostly due to reprocessing of the X-ray flux from the outer
accretion flows. Binary systems with low-mass companions to the neutron stars
or black holes are called Low-Mass X-ray Binaries (LMXBs).

A compact object can also accrete matter from a companion star that
does not fill its Roche lobe, if the latter star is losing mass in the
form of a stellar wind. For this process to result in a compact star
that is a bright X-ray source, the companion star has to be massive
($\ge 10$~$M_\odot$) in order to drive a strong wind. In this
configuration, the optical luminosity of the companion star dominates
the total emission from the system and the rate of mass transfer is
determined by the strength and speed \mk of the wind and the orbital
separation. Such systems are called High-Mass X-ray Binaries
(HMXBs).

\setcounter{figure}{0}
\begin{figure*}
 \centerline{
 \psfig{file=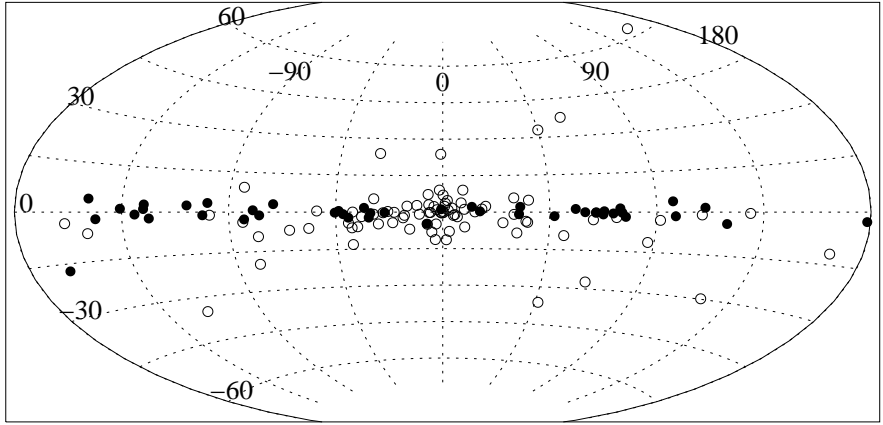,angle=0,width=12.truecm}}
 \caption{Distribution of Low-Mass X-ray Binaries (open symbols) and
 High-Mass X-ray Binaries (filled symbols) in galactic coordinates 
 (Grimm, Gilfanov \& Sunyaev 2002).}
 \label{fig:gal_distr}
\end{figure*}

The large difference in the companion masses between low- and
high-mass X-ray binaries leads to a number of additional differences
between these two classes of systems. The lifetimes of HMXBs are
determined by the evolution of the high-mass companions and are short
($\sim 10^5-10^7$~yr), whereas the lifetimes of the LMXBs are
determined by the mass-transfer process and are longer ($\sim
10^7-10^9$~yr). For this reason, HMXBs are distributed along the
galactic plane, as young stellar populations do, whereas \mk LMXBs are
\mk found mostly towards the galactic center and in globular clusters
(Fig.~\ref{fig:gal_distr}). Moreover, because neutron stars in HMXBs
accrete for a relatively short period of time, their magnetic fields
do not evolve away from their high birth values, and hence these
neutron stars appear mostly as accretion-powered pulsars. On the other
hand, the prolonged phase of accretion onto neutron stars in LMXBs is
believed to be responsible for the suppression of the stellar fields
and the absence of periodic pulsations in all but a handful of them.

Finally, in LMXBs, the very small sizes of the companion stars can
lead to a number of interesting configurations in systems that are
viewed nearly edge on. For example, in the Accretion Disk Corona (ADC)
sources, the X-rays from the central objects are scattered towards the
observer by electrons in a hot corona that has a size larger than that
of the companion, e.g., smoothing out the lightcurves of the X-ray eclipses
(White \& Holt 1982). On the other hand, in the so-called dippers,
the shallow X-ray eclipses may not be caused by the companion stars but
rather by the stream of mass transfer from the companion stars to the
accretion disks (see, e.g., White \& Swank 1982).

Overall, there are believed to be only a few hundred accreting
high-mass and low-mass X-ray binaries in the whole
galaxy. Consequently, these binaries are extremely rare among stellar
systems. This is in accord with the large number of improbable
evolutionary steps a primordial binary needs to follow in order to
become an X-ray source with an accreting compact object.  Indeed, the
progenitors of the compact objects are believed to be too large to fit
in the tight orbits of most X-ray binaries. Moreover, the supernova
explosions that precede the formation of the compact objects may
disrupt most systems at the phase prior to the formation of the X-ray
binary. The resolutions to these and other puzzles on the formation
and evolution of X-ray binaries involve exotic and poorly understood
binary-evolution processes such as common-envelope evolution of binary
stars (Taam
\& Sandquist 2000), asymmetric supernova explosions that impart recoil
velocities to the newborn compact objects, and two- and three-star
interactions in the dense stellar fields of globular clusters (see
\S8) . The processes that lead to the formation and evolution of
X-ray binaries are reviewed in detail in \S16.

\subsection{Accretion onto Compact Objects}

An X-ray binary is formed when either the companion star transfers
matter onto the compact object through the inner Lagrangian point or
the compact object captures mass from the wind of the companion star.
In both cases, the fate of the transferred mass depends on the amount
of angular momentum it possesses, on the physical processes by which it
looses angular momentum, and, most importantly, on the radiation
processes by which it cools (see Frank, King \& Raine 2002 for a
comprehensive review of this subject).

Beginning in the early 1970's and for the next two decades, most of
the modeling effort of accretion flows onto neutron stars and black
holes was based on two restrictive assumptions. First, accretion flows
were assumed to be loosing angular momentum at high rates because of
an unspecified process, 
% it's viscosity! \mk
with the effective kinematic viscosity
typically taken to be proportional to the pressure (see, e.g., Shakura
\& Sunyaev 1973; these solutions are often called $\alpha$-disks,
named after the constant of proportionality). Second, radiation
processes were assumed to be very efficient, so that the resulting
accretion flows were relatively cool, in the form of geometrically
thin accretion disks. The first of these assumptions stemmed from
calculations that showed the inefficiency of microscopic viscosity to
account for the high inferred rates of mass accretion in the observed
sources (see Pringle 1981 for a review). The second assumption, on the
other hand, was relaxed in a number of studies (e.g., Shapiro,
Lightman \& Eardley 1976) but the resulting solutions were shown to
be unstable (e.g., Piran 1978).

During the last decade, theoretical models of accretion flows onto
compact objects became increasingly more sophisticated and diverse
because of two major developments. First was the identification of a
magnetohydrodynamic instability in differentially rotating flows (the
magneto-rotational instability, or MRI; Balbus \& Hawley 1991, 1998),
which allows seed magnetic fields of infinitesimal strength in the
flow to get enhanced and tangled. This was shown to lead to a
fully-developed magnetohydrodynamic turbulence and provide an
efficient mechanism of angular momentum transport, as envisioned in
the earlier empirical models (Balbus \& Papaloizou 1999).  Studies of
the non-linear development of the instability, its level of saturation
(e.g., Sano et al.\ 2004), as well as of its effect on the overall
properties of accretion disks (e.g., Hawley \& Krolik 2001, 2002;
Armitage et al.\ 2001; McKinney \& Gammie 2002) require large-scale
numerical simulations and are all subjects of intense research
efforts.

\begin{figure*}
 \centerline{ \psfig{file=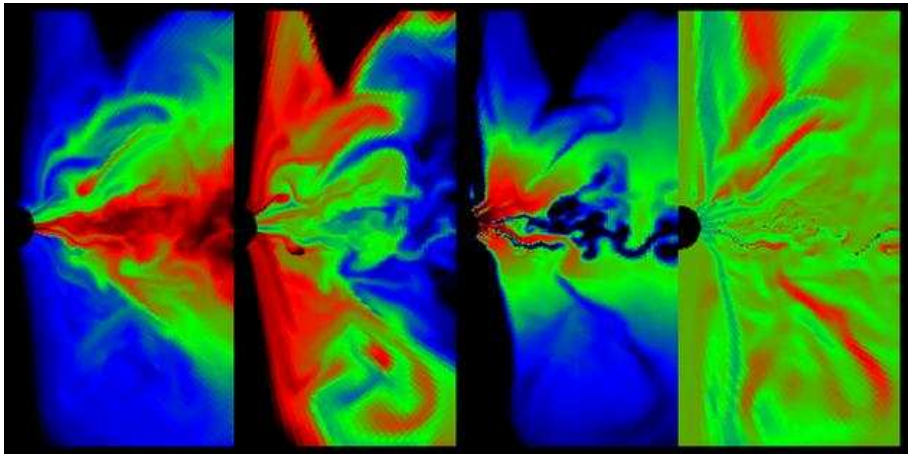,angle=0,width=12.truecm}}
 \caption{Results of a numerical simulation of a magneto-hydrodynamic
 accretion flow onto a black hole. The panels show the logarithm of
 the density, specific entropy, square of the toroidal magnetic field,
 and the $r-\phi$ component of the Maxwell stress tensor (Stone \&
 Pringle 2001).}  \label{fig:convection}
\end{figure*}

The second development is related to the discovery of new stable
% really? \mk
solutions to the hydrodynamic equations that describe radiatively
inefficient accretion flows (e.g., Narayan \& Yi 1994).  In these
solutions, the electrons and ions have different and high
temperatures, the accretion flows are geometrically thick, and most of
their potential energy is not radiated away but is rather advected
towards the compact objects; these are the so-called
Advection-Dominated Accretion Flows (ADAFs). Besides being interesting
new theoretical solutions to the hydrodynamics equations,
advection-dominated flows provided a framework within which the
anomalously low efficiency of accretion onto several black-holes in
centers of galaxies (Narayan et al.\ 1995) and in the quiescent states
of X-ray transients (Narayan et al.\ 1996) can be understood. 

Recently, a number of basic properties of these advection-dominated
solutions have been scrutinized.  The basic assumption that electrons
and ions are coupled inefficiently, mostly due to Coulomb scattering,
has been revised, taking into account the effects of magnetic fields
(e.g., Quataert \& Gruzinov 1999).  Advection-dominated flows were
shown to be capable of launching strong outflows (Advection-Dominated
Inflow/Outflow Solutions or ADIOS; Blandford \& Begelman
1999). Moreover, for a wide area of the parameter space, numerical
(Stone et al.\ 1999; Igumenshchev et al.\ 2000) and analytical studies
(Narayan et al.\ 2000) showed that the solutions are convectively
unstable (Convection-Dominated Accretion Flows or CDAFs). Finally, the
effects of the magneto-rotational instability on the properties of
radiatively inefficient flows have also been investigated recently
both in the Newtonian (see, Fig.~\ref{fig:convection}; Stone \&
Pringle 2001; Igumenshchev et al.\ 2003) and in the general relativistic
regimes (DeVilliers et al.\ 2003; Gammie et al.\ 2003).

The final ingredient in the models of accretion flows onto compact
objects is the interaction of the flows with the objects
themselves. This is the region in the accretion flows where most of
the high-energy radiation is produced and hence is the one that is
probed by observations with X-ray and $\gamma$-ray
telescopes. Clearly, the interaction depends on whether the compact
object is a black hole or a neutron star, and in the latter case, on
whether the neutron star is strongly or weakly magnetic. The main
observational manifestation of these differences is the presence or
absence of pulsations in the X-ray lightcurves of the systems, which
reflects the strength of the magnetic fields of the central objects.
In the rest \mk of this chapter \mk the observational properties of the
pulsating and non-pulsating X-ray binaries, as well as the current
efforts for their theoretical modeling, will be reviewed.

\section{Pulsating neutron stars}

Neutron stars possess some of the strongest magnetic fields observed in
nature. The origin of these magnetic fields is only poorly understood,
mostly due to our inability to observe directly
% really ? \mk
the magnetic fields of the cores of pre-supernova stars, which
collapse to form the neutron stars, and model their
amplification. However, observations of isolated radio pulsars
(\S7\mk) and magnetars (\S14\mk) provide strong evidence that
neutron-star magnetic fields range between $\sim 10^{8}~$G and $\sim
10^{15}$~G.

When a strongly magnetic neutron star accretes plasma \mk from a companion
star or the interstellar medium, its magnetic field becomes
dynamically important close to the stellar surface and determines the
properties of the accretion flow. The radius at which the effects of
the magnetic field dominate all others is called the Alfv\'en radius
and its precise definition depends on the mode of accretion (i.e.,
thin-disk vs.\ quasi-radial), the topology of the magnetic field
(i.e., dipolar vs.\ multipolar), etc. For thin-disk accretion onto a
neutron star, the Alfv\'en radius is defined as the radius at which
magnetic stresses remove efficiently the angular momentum of the
accreting material (see Ghosh \& Lamb 1991 and references
therein). For a surface magnetic field strength of $10^{12}$~G and a
mass accretion rate comparable to the Eddington critical rate, the
Alfv\'en radius is of order $\sim 100$ neutron-star radii.
% plse define magnetosphere and 'ic radius here \mk

The fate of the accreting material after it interacts with the stellar
magnetic field near the Alfv\'en radius depends on the spin frequency
of the neutron star. If the stellar spin frequency is smaller than the
orbital frequency of matter at the interaction radius, then the
accreting material is forced into corotation with the star and is
channeled along field lines onto the magnetic poles. As the neutron
star spins and the observer sees a different aspect of the hotter
magnetic poles, the X-ray flux received is modulated at the stellar
spin frequency and an accretion-powered pulsar is produced. On the
other hand, if the stellar spin frequency is larger than the orbital
frequency of matter at the interaction radius, then the material
cannot overcome the centrifugal barrier in order to accrete onto the
star. The fate of matter in this case is presently unknown,
% can't it accrete at all? \mk
but it is often assumed that matter eventually escapes the neutron
star in the form of a wind. Magnetic neutron stars in this
configuration are often said to be in the propeller regime (after
Illarionov \& Sunyaev 1975).

The neutron star itself also reacts differently to the accretion of
matter depending on its magnetic field strength, its spin frequency,
and the mass accretion rate. Magnetic field lines rotate at the spin
frequency of the star and couple the stellar surface to the accreting
material. As a result, they transfer angular momentum from the
accreting material to the neutron star, if the former is spinning
faster than the latter or from the neutron star to the accreting
material, in the opposite situation. Both situations occur
simultaneously in an accreting system, since the orbital frequency of
matter decreases with increasing radius. The overall effect is a net
torque on the neutron star, which can be either positive (spin-up) or
negative (spin-down). The magnitude of the torque on the star is
expected to increase with increasing mass accretion rate and with
increasing magnetic field strength (see Ghosh \& Lamb 1979; Ghosh \&
Lamb 1991). Clearly, for every magnetic field strength and mass
accretion rate, there is a critical spin frequency at which the net
torque on the star is zero. This frequency corresponds to an
equilibrium, towards which the neutron star evolves in its lifetime.
For a surface, dipolar magnetic field with a strength of $10^{12}$~G
and a mass accretion rate comparable to the Eddington critical rate,
the equilibrium spin frequency is of order of a few tenths of a Hertz
(Ghosh \& Lamb 1992).

Accretion-powered pulsars provide currently the best systems in which
the spin frequencies and magnetic field strengths of accreting neutron
stars can be studied. Two distinct classes of such pulsars are known:
pulsars with periods of order a second, which are found mostly in
high-mass X-ray binaries, and pulsars with millisecond spin periods,
which are found in binary systems with very short orbital periods (see 
\S 1.1.2).

\subsection{Classical (slow) accretion-powered pulsars}

The detection of coherent pulsations from an accreting X-ray source,
in 1971 (Giacconi et al.\ 1971), provided the strongest evidence, at
the time, that the compact objects in many of these sources were
neutron stars. Since then, accretion-powered pulsars with periods of
the order of one second or more have been studied extensively with
every X-ray satellite. In recent years, the long-term monitoring of
such pulsars with BATSE as well as the detailed spectral studies with
{\em RXTE\/} and {\em BeppoSAX\/} provided a unique look into the
properties of these systems, resolving some long-standing questions
and posing a number of new ones (see Bildsten et al.\ 1997 and Heindl
et al.\ 2004 for comprehensive reviews of the accretion-powered
pulsars discussed in this section).

\begin{figure*}
 \centerline{
 \psfig{file=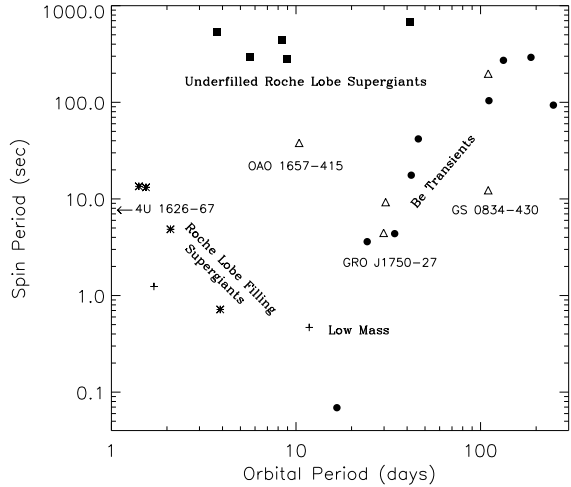,angle=0,width=8truecm}}
\caption{The spin and orbital periods of classical accretion-powered pulsars
(the Corbet diagram; after Bildsten et al.\ 1997).}
\label{fig:corbet_old}
\end{figure*}

The vast majority of slow accretion-powered pulsars are found in
high-mass X-ray binaries; only five of them (Her~X-1, 4U~1626$-$67,
GX~1$+$4, GRO~J1744$-$28, and 2A~1822$-$371) have low-mass
companions. Indeed, most low-mass X-ray binaries are old systems and
their prolonged phase of accretion is thought to have suppressed the
magnetic fields of the neutron stars and to have spun them up to
millisecond periods (see
\S\ref{sect:mspsr}). On the other hand, high-mass X-ray binaries
are younger systems and the neutron stars in them are expected to 
have magnetic fields that are dynamically important.

The properties of the high-mass binary systems in which slow pulsars
reside can be described more easily on the diagram that correlates
their spin to their orbital periods (the Corbet diagram;
Fig.~\ref{fig:corbet_old}). About half of the slow pulsars are
orbiting main-sequence Be stars, whereas the remaining pulsars are
orbiting evolved OB supergiants. The systems with Be companions are
generally eccentric transient systems, in which the companion stars
are not filling their Roche lobes and the pulsars become detectable
during periastron passages (\S5\mk). The properties of the systems with
supergiant companions depend on whether these stars fill their Roche
lobes or not. If they do, matter is transfered onto the neutron stars
via the inner Lagrangian point of their binary potential possessing
significant angular momentum and forming a geometrically thin
accretion disk.  
% not stable as said before \mk
On the other hand, for companion stars that do not 
fill their Roche lobes, mass lost via a radiation-driven wind is
captured by the neutron star at rates that are typically lower than
the disk-fed systems.

\subsubsection{Spin-period evolution}

The mode of mass transfer onto the neutron stars, which depends on the
properties of the binary \mk systems, also determines the spin evolution
of the neutron stars. The long-term aspect of this dependence is
clearly visible in Fig.~\ref{fig:corbet_old}. The systems with
Roche-lobe filling supergiants have short spin periods that are
anticorrelated with the orbital periods; the systems with underfilling
supergiants have long spin periods that do not show any correlation with 
orbital periods; and the Be transient systems have long orbital periods
which are positively correlated to the orbital periods. These correlations
are believed to depend strongly on the mode and efficiency of mass
transfer from the companion stars to the neutron stars but are only
poorly understood (see, e.g., Waters \& van Kerkwijk 1989). 

A clear look into the short-term dependence of the spin periods of
neutron stars on the properties of the accretion flows was made
possible \mk because of the intense monitoring of several
accretion-powered pulsars with the BATSE experiment onboard {\em
CGRO}. Contrary to earlier results, the measurements with BATSE
revealed that transient and persistent sources show two different
types of spin-period evolution (Bildsten et al.\ 1997).

\begin{figure*}
 \centerline{
 \psfig{file=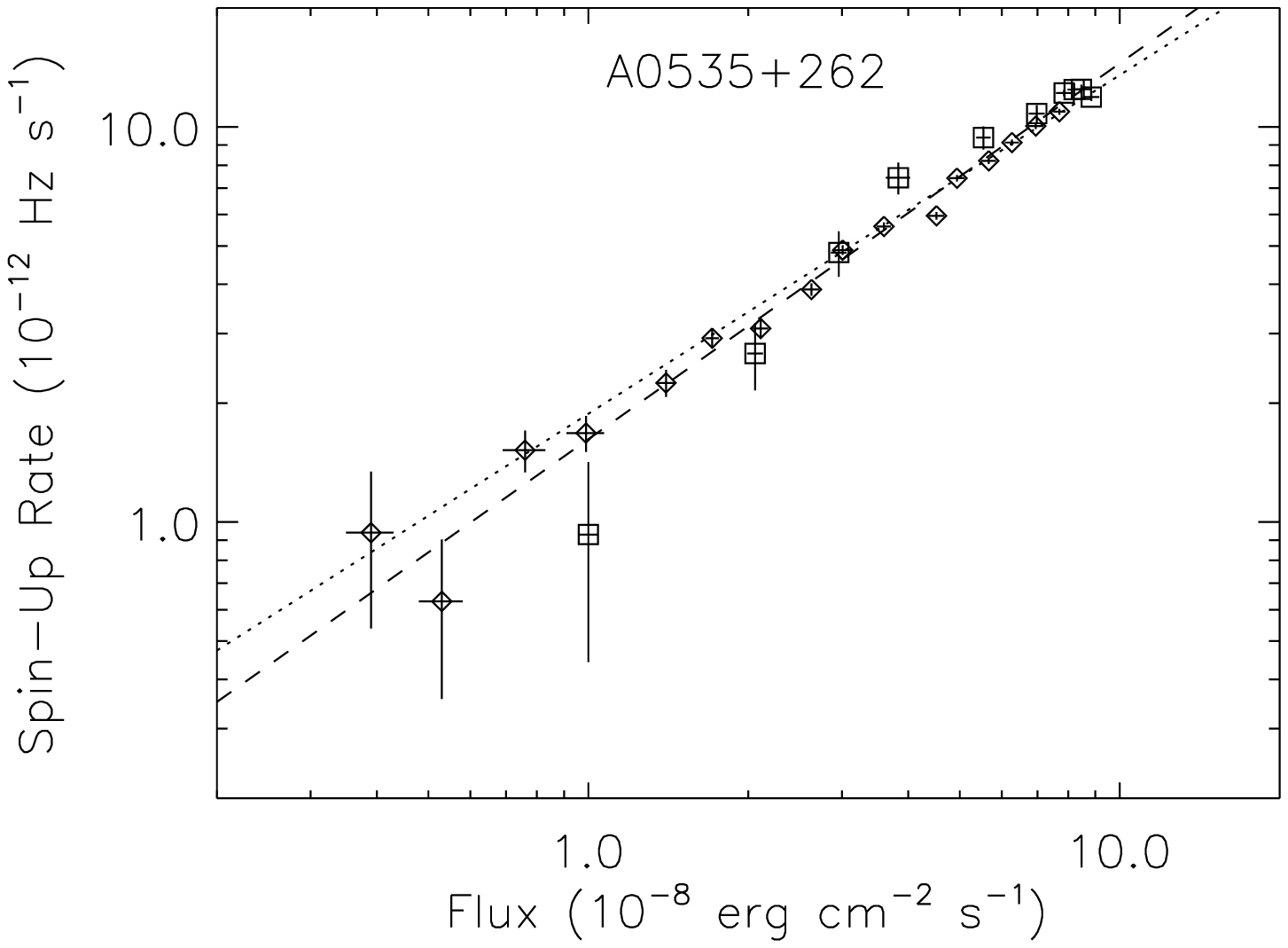,angle=0,height=4.5truecm,width=6truecm}
 \psfig{file=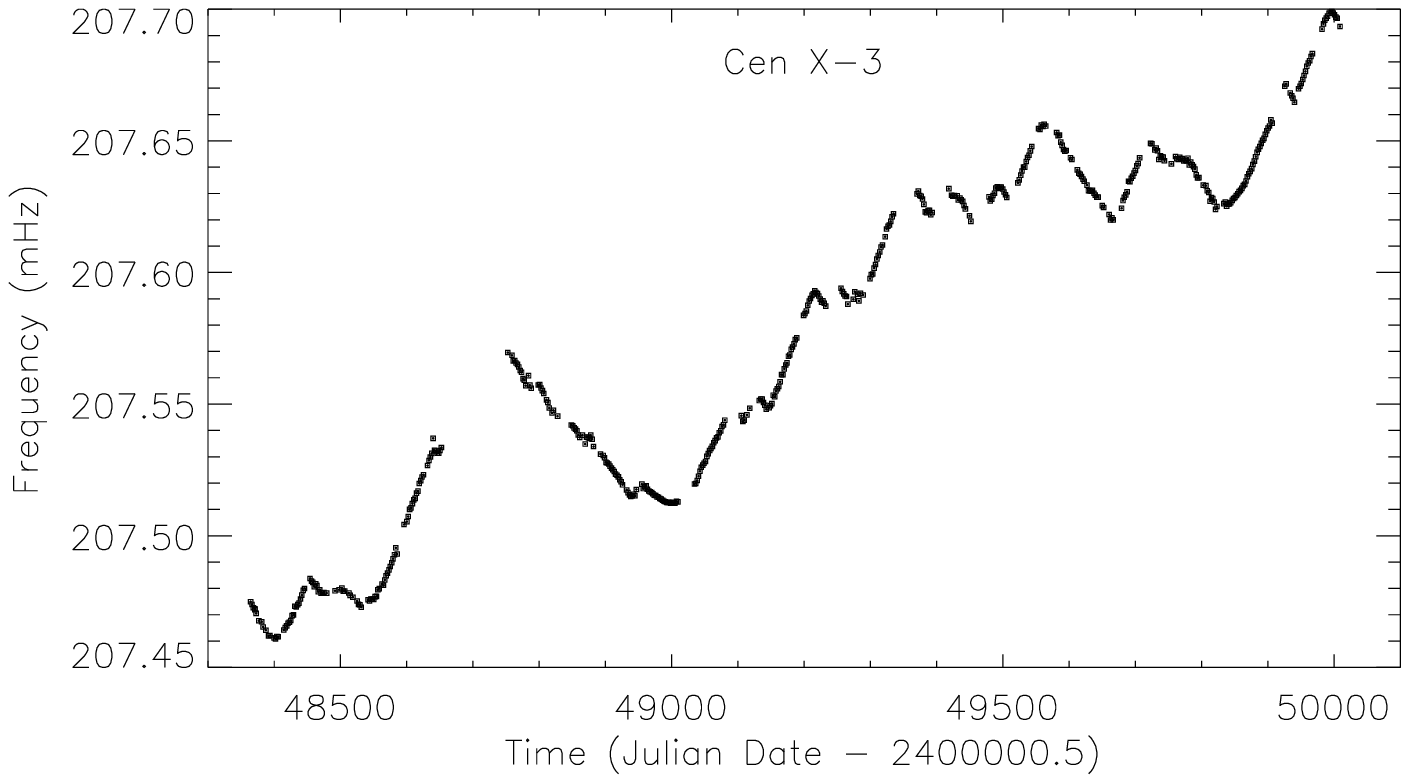,angle=0,height=4.5truecm,width=5.5truecm}}
\caption{{\em (Left)\/} The dependence of the spin-up rate on the pulsed
flux for the source A~0535$+$262, as observed by BATSE; the dashed and
dotted curves show the best-fit line and the theoretical prediction,
respectively (see text).  {\em (Right)\/} The evolution of the spin
frequency of Cen~X-3 as observed by BATSE (Bildsten et al.\ 1997).}
\label{fig:acc_psrs}
\end{figure*}

\begin{table}
\caption{Quasi-Periodic Oscillations in Accretion-Powered Pulsars$^a$}
\label{tab:psr_qpo}
\begin{tabular}{lcc}
\hline\hline
{Source} &  {Spin Frequency (mHz)} & {QPO frequency (mHz)} \\
\hline
4U~1907$+$09 & 2.27  & 55 \\
XTE~J1858$+$034 & 4.5 & 111\\
A~0535$+$26  & 9.71  & 27--72 \\
EXO~2030$+$375 & 24 & 187--213\\
LMC~X-4      & 74    & 0.65--1.35, 2--20 \\
4U~1626$-$67 & 130   & 1,48\\
Cen~X-3      & 207   & 35 \\
V~0332$+$53  & 229   & 51\\
4U~0115$+$63 & 277   & 2, 62 \\
Her~X-1      & 807.9 & 8, 12, 43 \\
SMC~X-1      & 1410  & 60? \\
GRO~1744$-$28$^b$ & 2140 & 40000\\  
\hline
\hline
\end{tabular}

{\footnotesize $^a$~compilation after Shirakawa \& Lai 2002; $^b$
Zhang et al.\ 1996}
\end{table}

Transient accretion-powered pulsars in outburst show a positive
dependence of the accretion torque (as measured by the spin-up rate)
on the inferred accretion luminosity (Fig.~\ref{fig:acc_psrs}; left
panel). This is consistent with the simple model of disk-magnetosphere
interaction (e.g., Ghosh \& Lamb 1992), in which, as the accretion
rate increases, the rate of angular momentum transfer from the
accretion flow to the neutron star increases. At the limit of very low
mass accretion rate, the neutron stars are expected to spin down,
because the magnetic field lines that couple to the outer, slower
accretion flow remove spin angular momentum from the neutron
star. Such spin-down episodes have not been detected by BATSE,
although evidence for spin down in the pulsar EXO~2030$+$375 has been
previously reported based on {\em EXOSAT\/} data (Parmar et al.\
1989). The very low fluxes down to which the transient sources
continue to spin up place strong constraints on the relative
importance of angular momentum transfer between the accretion disk and
the neutron star via anchored magnetic field lines.

In sharp contrast to the transient sources, persistent disk-fed
pulsars show a bimodal behavior in their accretion torques (Bildsten
et al.\ 1997; see also Fig.~\ref{fig:acc_psrs}; right panel). Episodes
of spin-up and spin-down of approximately equal accretion torques
alternate at timescales that vary from $\sim 10$~days (e.g., in
Cen~X-3) to $\ge 10$~yr (e.g., in GX~1$+$4). The transition between
spin-up and spin-down is rapid ($\le$~a few days) and cannot be
resolved with BATSE measurements.  Current models of the
disk-magnetosphere interaction in accretion-powered pulsars can
account for the observed bimodal torques only if one of the physical
properties of the accretion flow is also assumed to show a bimodal
behavior. Such assumptions include a bimodal distribution of the mass
transfer rate onto the pulsar, or a bimodal dependence on accretion
rate of the orientation of the disk (Nelson et al.\ 1998; van Kerkwijk
et al.\ 1998), of the orbital angular velocity of the accreting gas
(Yi \& Wheeler 1998), or of the strength and orientation of any
magnetic field produced in the disk (Torkelsson 1998). Alternatively,
for any given mass accretion rate onto the neutron star, two
equilibrium solutions may be possible, one in which the star is
spinning up and one in which it is spinning down (Lovelace et al.\
1999). It is not clear at this point which, if any, of these
alternatives is responsible for the observed torque reversals in
accreting neutron stars and this remains one of the puzzles of the
BATSE monitoring of slow accretion-powered pulsars.

\subsubsection{Quasi-periodic oscillations}

\begin{figure*}
 \centerline{
 \psfig{file=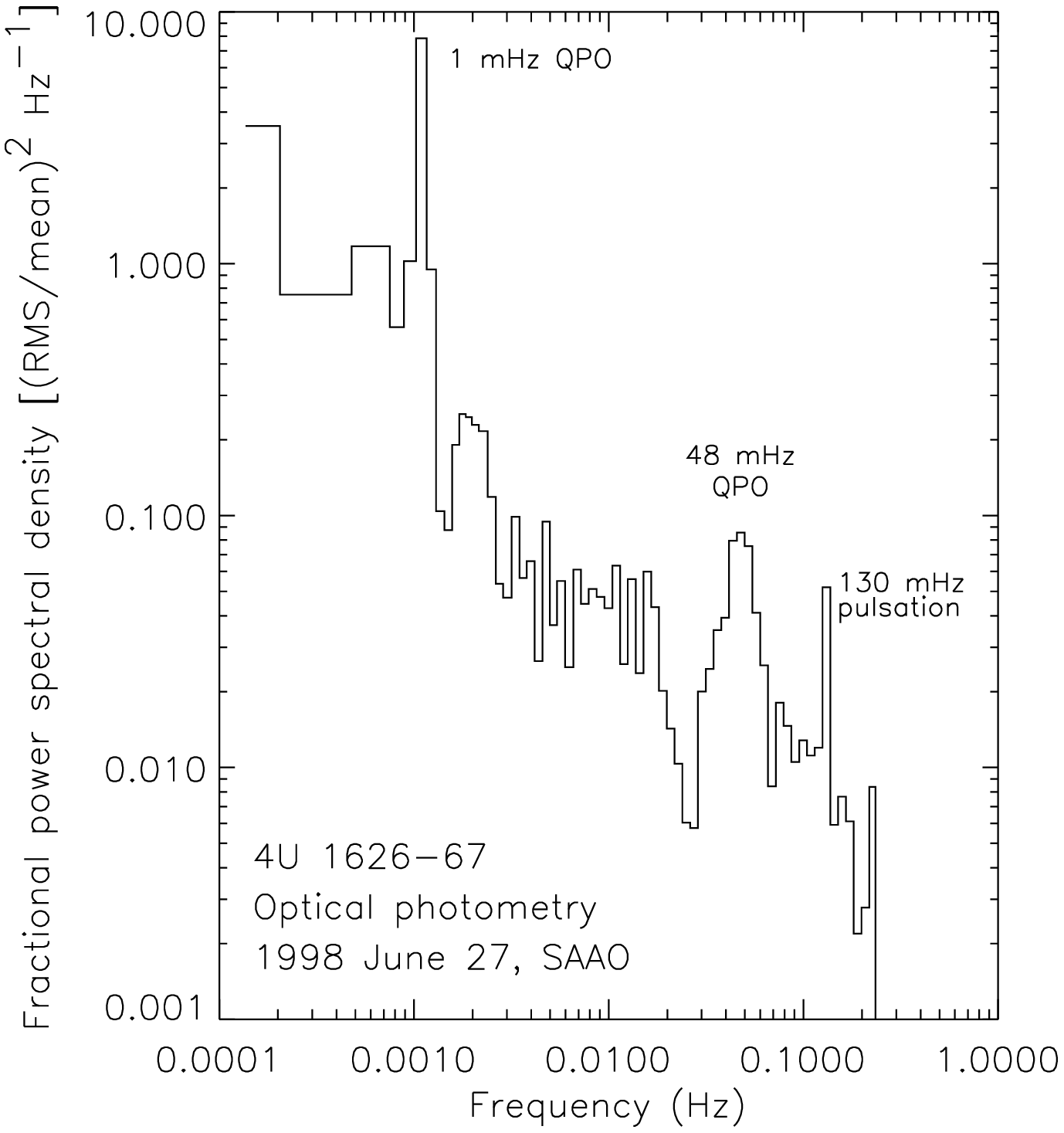,angle=0,width=6truecm}
 \psfig{file=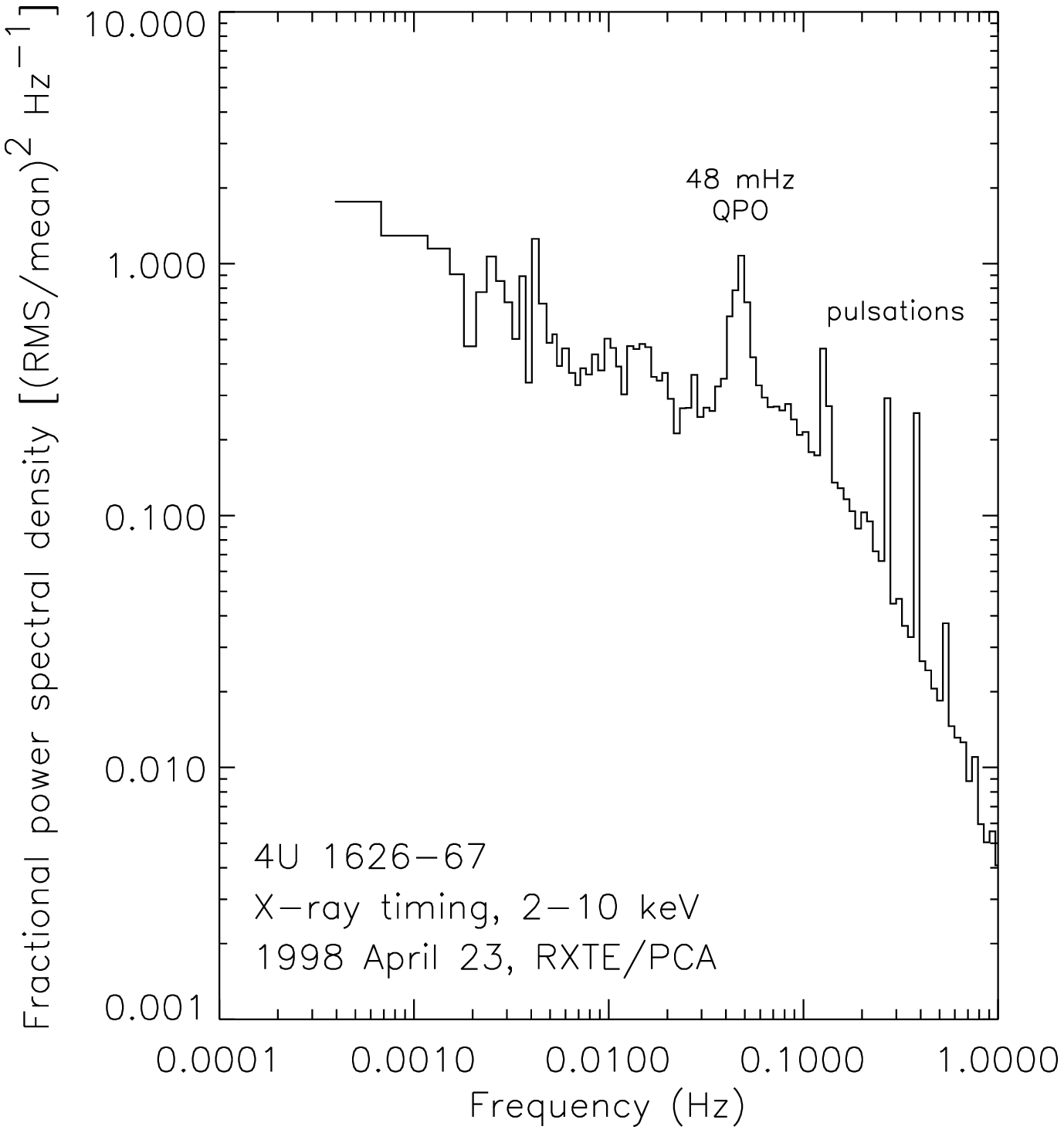,angle=0,width=6truecm}}
\caption{The optical (left) and X-ray (right) power spectrum of the
accretion-powered pulsar 4U~1626$-$67 showing mHz quasi-periodic
oscillations (Chakrabarty et al.\ 2001).}
\label{fig:qpo_1626}
\end{figure*}

In several accretion-powered pulsars, the power-density spectra in
X-rays or in longer wavelengths show a number of quasi-periodic
oscillations, in addition to the period of the pulsars
(Table~\ref{tab:psr_qpo}; Fig.~\ref{fig:qpo_1626}; \S2.11\mk). The frequencies of
these oscillations range from $\simeq 1$~mHz to $\simeq 40$~Hz and
they can be from $\sim 100$ times smaller to $\sim 100$ larger than
the pulsar spin frequencies.

The frequency of the fast oscillations in the transient pulsar
EXO~2030$+$375 was \mk found to be in good agreement with beat-frequency
models (Finger et al.\ 1998), in which oscillations occur at the beat
frequency between the orbital frequency of matter at the Alfv\'en
radius and the stellar spin frequency (Alpar \& Shaham 1995). On the
other hand, the low-frequency oscillations observed in 4U~1626$-$67
appear also as asymmetric sidebands to the pulse period (Kommers et
al.\ 1998) and are probably related to a low-frequency modulation of
the accretion flow, possibly due to the presence of a precessing disk
warp (Shirakawa \& Lai 2002).  It is unclear at this point whether all
of these quasi-periodic oscillations are related to the same
phenomenon or not and what is their physical origin.

\subsubsection{Cyclotron lines}

The X-ray spectra of accretion-powered pulsars are typically described
in terms of relatively flat power-laws with exponential cut-offs at
energies $\ge 10$~keV. These continuum spectra are believed to be the
result of upscattering of soft photons by the hot electrons in the
accretion columns above the magnetic polar caps (Meszaros 1992). For
neutron-star magnetic field strengths of $\simeq 10^{12}$~G, the
cyclotron energy on the stellar surface is $\simeq 11.6$~keV and the
continuum spectra are expected to show evidence for harmonically
related ``cyclotron resonance scattering features'' (or simply
cyclotron lines) in the X-rays. Observation of such features was
anticipated from the early days of X-ray astronomy and \mk expected to
lead to direct measurements of the magnetic field strengths of
accreting neutron stars (e.g., Trumper et al.\ 1978).

\begin{figure*}
 \centerline{
 \psfig{file=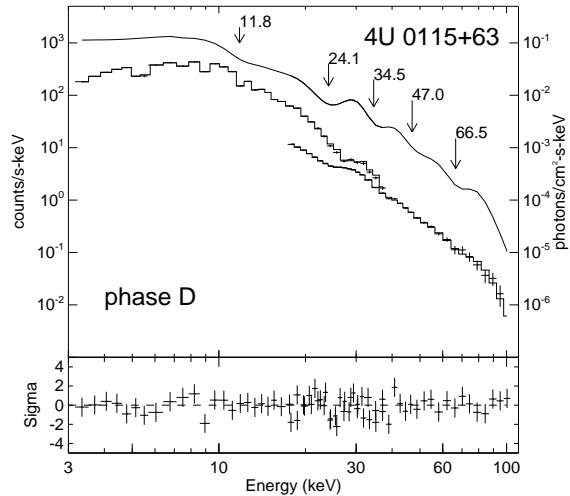,angle=0,width=8truecm}} 
\caption{The observed (histogram) and model spectrum (solid line) of the
accretion-powered pulsar 4U~0115$+$63 showing evidence for cyclotron lines
with as many as four overtones (Heindl et al.\ 1999).}
\label{fig:cycl}
\end{figure*}

\begin{table}
\caption{Energies of Cyclotron Lines in Accretion-Powered Pulsars$^{a}$}
\label{tab:cycl}
\begin{tabular}{lccc}
\hline\hline
{Source} &  {Spin Period (s)} & {Energy (keV)} & {Harmonics}\\
\hline
4U~0115$+$63 & 3.61 & 12 & Yes\\
4U~1907$+$09 & 438  & 18 & Yes\\
4U~1538$-$52 & 530  & 20 & Yes\\
Vela~X-1     & 283  & 25 & Yes\\
V~0332$+$53  & 4.37 & 27 & \\
Cep~X-4      & 66.2 & 28 & \\
Cen~X-3      & 4.82 & 28.5 & \\
4U~0352$+$309  & 835  & 29 & \\
XTE~J1946$+$274 & 15.8 & 36 & \\
MX~0656$-$072 & 160.7 & 36 & \\
4U~1626$-$67 & 7.66 & 37 & \\
GX~301$-$2   & 681  & 37 & \\
Her~X-1      & 1.24 & 41 & \\
A~0535$+$26  & 105  & 50 or 110 & \\  
\hline
\hline
\end{tabular}
% nr of harmonics? \mk

{\footnotesize $^a$~after Heindl et al.\ 2004}
\end{table}

The broadband spectral capabilities of {\em RXTE\/} and {\em
BeppoSAX\/} made possible the unequivocal detection of harmonically
related features in four accretion-powered pulsars, as well as the
detection of single features in ten more sources (Fig.~\ref{fig:cycl}
and Table~\ref{tab:cycl}). The widths of the resonance features appear
to be correlated to the energies of the continuum cut-offs and to be
proportional to their central energies and to the inferred scattering
depths (Coburn et al.\ 2002). The central energies of these features
provided direct measurements of the surface magnetic fields of
accreting \mk pulsars, which can be used in constraining the models of
disk-magnetosphere interaction.  Future observations of the
pulse-phase dependence of the scattering features is also expected to
provide more detailed constraints on the geometries of the accretion
columns in slow accretion-powered pulsars.
% not yet done w/ SAX? \mk

\subsection{Millisecond accretion-powered pulsars}
\label{sect:mspsr}

The presence of accretion-powered millisecond pulsars in low-mass
X-ray binaries had been predicted 17 years before such sources were
eventually discovered. Millisecond rotation-powered pulsars were
thought to acquire their low ($\simeq 10^8$~G) magnetic fields and
fast spin frequencies while accreting mass at high rates in low-mass
X-ray binaries (Alpar et al.\ 1982; Radhakrishnan \& Shrinivasan
1982). These millisecond radio pulsars were most often found in
binaries with evolved, low-mass white dwarf companions (Bhattacharya
\& van den Heuvel 1991), which were thought to be the descendents of
LMXBs, and their birthrates were similar (albeit with systematic
differences) to those of LMXBs (see, e.g., Kulkarni \& Narayan 1988;
Lorimer 1995 and references therein). Moreover, circumstantial
evidence based on the bursting behavior (Lewin, van Paradijs, \& Taam
1996), rapid variability (Alpar \& Shaham 1985; Ghosh \& Lamb 1991),
and X-ray spectra (Psaltis \& Lamb 1998) of the compact objects in
bright LMXBs strongly supported their identification with
weakly-magnetic, rapidly spinning neutron stars. However, despite
intense searches (e.g., Vaughan et al.\ 1994), periodic pulsations
could not be detected from any LMXB, making this the holy grail of
X-ray binary astrophysics in the pre-{\em RXTE\/} era.

\begin{figure*}
 \centerline{
 \psfig{file=1808pds.ps,angle=-90,height=4truecm,width=5.5truecm}
 \psfig{file=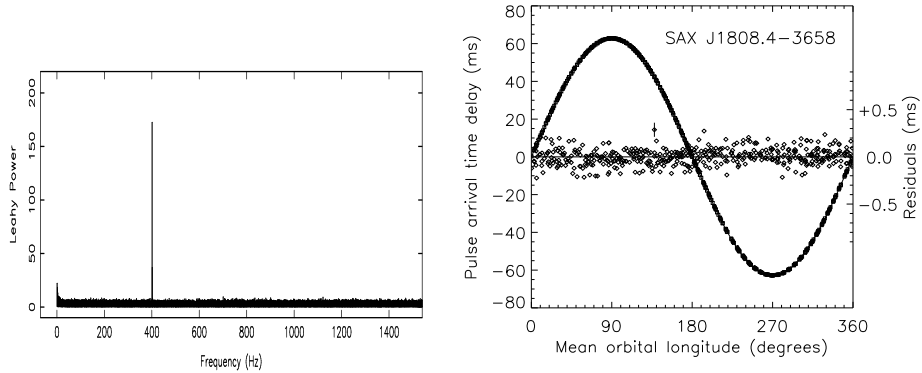,angle=0,width=6truecm}} 
\caption{The power-density spectrum of the X-ray flux from SAX~J1808.4$-$3658
 showing clearly the $\sim 401$~Hz pulsation (left; Wijnands \& van der
 Klis 1998) and the pulse-time residuals caused by its 2~hr orbit
 (right; Chakrabarty \& Morgan 1998).}  
\label{fig:1808}
\end{figure*}

The discovery, with {\em RXTE}, of highly coherent pulsations in the
X-ray fluxes of LMXBs during thermonuclear X-ray bursts (Strohmayer et
al.\ 1996; see also Chapter~3 in this volume) provided the then \mk strongest
evidence for the presence of neutron stars with millisecond spin
periods in LMXBs. However, the first bona fide millisecond, accretion
powered pulsar was discovered only \mk in 1998, in a
transient ultracompact binary (see Fig.~\ref{fig:1808}). Since then,
four additional millisecond pulsars were discovered in very similar
transient binaries (see Table~\ref{tab:mspsrs}).

\begin{table}
\caption{Observed Properties of Millisecond Accreting Pulsars$^{a}$}
\label{tab:mspsrs}
\begin{tabular}{lcccc}
\hline\hline
{Source} & {$f_{\rm s}$} (Hz) & {$P_{\rm orb}$} (m) & {$a$ (lt-ms)} & {$f$~$(M_\odot)$}\\
\hline
SAX~J1808.4$-$3658$^b$ & 401.0 & 120.9 & 62.809 
    & $3.78\times 10^{-5}$ \\
XTE~J0929$-$3314$^c$ & 185.1 & 43.6 & 6.290 
    & $2.7\times 10^{-7}$ \\
XTE~J1751$-$305$^d$ & 435.3 & 42.4 & 10.1134 
    & $1.278\times10^{-6}$ \\
XTE~J1807$-$294$^e$ & 190.6 & 40.1 & 4.80 & $1.54\times 10^{-7}$\\
XTE~J1814$-$338$^f$ & 314.3 & 256.5 & 390.3 & $2.016\times 10^{-3}$\\
\hline
\hline
\end{tabular}

{\footnotesize $^a$~Spin frequency, orbital period, projected
semi-major axis, and mass function; $^b$~Wijnands \& van der Klis 1998
and Chakrabarty \& Morgan 1998; $^c$~Galloway et al.\ 2002;
$^d$~Markwardt et al.\ 2002; $^e$~Markwardt et al.\ 2003; Markwardt,
priv.\ comm.; $^f$~Markwardt \& Swank 2003; Markwardt, priv.\ comm.}
\end{table}

\begin{figure*}
 \centerline{
 \psfig{file=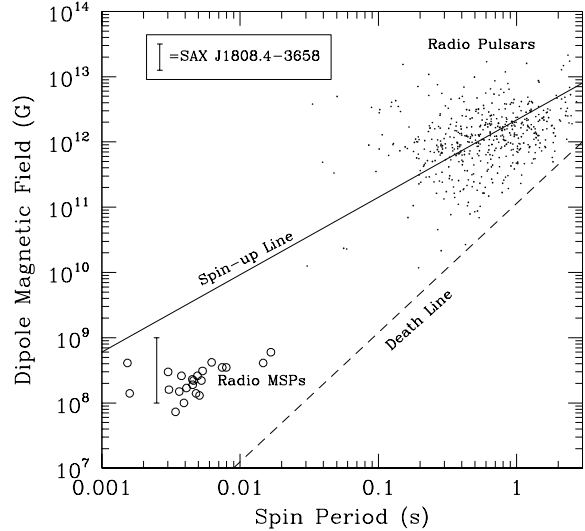,angle=0,width=8truecm}} 
\caption{Inferred strengths of the dipole magnetic fields of
slow radio pulsars (dots), millisecond radio pulsars (open circles),
and of the accretion-powered millisecond pulsar SAX~J1808.4$-$3658
(Psaltis \& Chakrabarty 1999).}
\label{fig:bp}
\end{figure*}

The spin periods and inferred magnetic fields of these five pulsars
are indeed consistent with the prediction that such systems are the
progenitors of the rotation-powered millisecond pulsars observed in
radio wavelengths (see, e.g., Fig.~\ref{fig:bp}; Psaltis \&
Chakrabarty 1998). However, the binary systems \mk which all five
pulsars belong to have a number of unusual characteristics. First,
they are all ultracompact as inferred from their small orbital periods
($\le 2$~hr) and projected semi-major axes ($\le 62$~lt-ms). Moreover,
the companions to the neutron stars have masses of only a few
hundredths of the solar mass and are, most probably, the remnants of
white dwarfs that have lost most of their mass. 
% plse provide a ref for this \mk
Finally, the inferred
long-term averages of the mass accretion rates onto these neutron
stars, as well as their maximum luminosities during outbursts are
among the lowest in the known LMXB population (see, e.g., Chakrabarty
\& Morgan 1998). 

Although they resolved one of the long-standing puzzles in X-ray
binary astrophysics, these five millisecond pulsars have posed two
very important questions: First, why do these particular neutron stars in
these special binaries appear as X-ray pulsars, whereas the compact
objects in the other LMXBs do not; and second, why does the spin-up of
these systems stop \mk short of the sub-millisecond periods that most
neutron-star models allow (see, e.g., Glendenning 2003 and references
therein). No satisfactory answer to the first question has been found
to date, whereas three equally exciting answers to the second question
appear plausible.

Whether an accreting neutron star appears as an X-ray pulsar depends
mostly on three factors: its magnetic field, the mass accretion rate,
and the relative orientation of the binary, spin, and magnetic axes to
the direction of the observer. For example, unless the binary axis
makes a relatively small angle with the direction of the observer, the
accretion flow may block the direct viewing of the polar caps and
hence the modulation of the X-ray flux at the stellar spin
frequency. Motivated by the very small measured values of the mass
function of SAX~J1808.4$-$3658 (see, Table~\ref{tab:mspsrs}), which
suggest a priori small inclination angles, Psaltis \& Chakrabarty
(1998) suggested that a favorable viewing geometry is responsible for
the fact that this source appears as an X-ray pulsar. However, the
subsequent discovery of four additional pulsars in similar binary
systems (albeit with equally small mass functions), the discovery of a
small modulation of the X-ray flux at the orbital period of
SAX~J1808.4$-$3658 \mk 
(Chakrabarty \& Morgan 1998), as well as the optical properties of the
reprocessed radiation from the accretion disk (Wang et al.\ 2000) all
argue against such a favorable geometry.

Alternatively, if the Alfv\'en radius is smaller than the size of the
neutron star then the stellar magnetic field is nowhere dynamically
important and cannot channel the accretion flow preferentially onto
the magnetic poles. Cumming et al.\ (2001) argued that in the
accreting pulsars \mk the very low
rate of mass transfer onto the neutron stars is not sufficient to
``bury'' their magnetic fields and make them dynamically unimportant, in
contrast to the case of the other LMXBs, that are accreting at much
larger rates on average.  This is consistent with the peculiar
properties of the binaries in which all five millisecond accreting
pulsars reside. It is, however, hard to reconcile with the fact that
these five sources, besides being pulsars, have very similar
X-ray spectra (Gilfanov et al.\ 1998) and aperiodic variability
properties (Wijnands \& van der Klis 1998) with many non-pulsing
sources. 
% didn't Pouttanen or somebody claim the spectrum was exceptional?
% and the frequency-frequency correlations certainly are! see van
% straaten. also: why are not all weak transients pulsars? \mk

It appears that, in the five millisecond pulsars, the stellar magnetic
fields only introduce \mk a modulation of the X-ray flux at the stellar
spin frequency without altering significantly the other timing and
spectral properties of the systems, and hence their accretion
flows. 
% see above
Consistent with this fact, these five pulsars have spin
frequencies in the same range as the ones inferred in most \mk other,
non-pulsing LMXBs from the properties of burst oscillations
(Chakrabarty et al.\ 2003). Indeed, the spin frequencies of all these
sources are limited to be less than $\simeq 700$~Hz, which is also
similar to the spin frequency of the fastest known millisecond,
rotation-powered pulsar. The physics that sets this limit is unclear
at this point.

As the simplest explanation, general relativity and the equation of
state of neutron-star matter may not permit stable neutron stars with
spin frequencies faster than $\simeq 700$~Hz. 
% how does this differ from nr 3 option? \mk
Albeit reasonable, this
limit on the spin frequency would require neutron stars to be about
twice as large as predicted by most models and hence, if correct,
would point to a more exotic equation of state than any currently
discussed (Cook, Shapiro, \& Teukolsky 1994). Alternatively, if the
neutron stars in LMXBs spin near the magnetic spin equilibrium, there
may be a natural upper limit on spin frequency that magnetic accretion
can achieve. This would require a lower bound on the stellar magnetic
field, since a non-magnetic neutron star will always spin-up to the
maximum frequency allowed by its equation of state, and hence would
provide significant clues for the efficiency of magnetic field decay
in accreting neutron stars.

The most exciting alternative, however, is the possibility that
neutron stars spinning at rates faster than about $700$~Hz rapidly \mk
loose their spin angular momenta via emission of gravitational
radiation (Bildsten 1998). For this to happen, the distribution of
matter inside the neutron star must be non-axisymmetric since the
emission of gravitational radiation depends on the time derivative of
the mass quadrupole \mk of the star. Temperature anisotropies in the
surface layers of accreting neutron stars may result in anisotropies
in the crystallization of the material underneath that are sufficient
to account for the rapid loss of spin angular momentum (Bildsten
1998). Alternatively, excitation of non-radial modes in the neutron
star may provide such a time-dependent mass quadrapole (Andersson,
Kokkotas, \& Schutz 1998). If this is the reason why the spin-up of
accreting neutron stars stalls at a frequency well below the maximum
frequency allowed by their equation of state, then LMXBs may become
the first sources detected in the very near future by gravitational
wave observatories.

\section{Non-pulsing neutron stars and black holes}

The large majority of accreting compact objects show no evidence for
periodic pulsations in their persistent emission. In the case of
accreting black holes, this is a direct consequence of the presence of
the event horizon, which does not allow for any stable feature to be
anchored to the rotation of the compact object. In the case of neutron
stars, however, the absence of pulsations requires a rather weak
magnetic field ($\le 10^8$~G) so that the accretion flow is not
disrupted and channeled onto the magnetic poles.

\begin{figure*}
 \centerline{
 \psfig{file=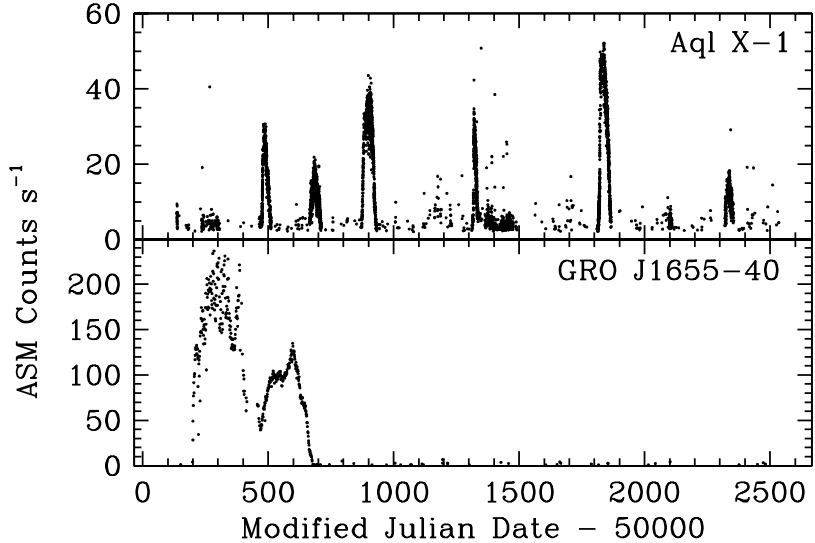,angle=0,width=11truecm}}
 \caption{Lightcurves of a neutron-star (Aql~X-1) and a black-hole
(GRO~J1655$-$40) transient source, as observed by the All Sky Monitor
on {\em RXTE}.}
 \label{fig:trans}
\end{figure*}

\subsection{Transient and Persistent sources}

Non-pulsing accreting compact objects appear both as persistent and as
transient sources. Members of the first class are observed at X-ray
fluxes that can be variable by up to several factors of two over
timescales ranging from milliseconds to months. Their distinguishing
characteristic, however, is the fact that they have been at
detectable flux levels for most of the history of X-ray astronomy.

Transient sources, on the other hand, are characterized by long
periods of inactivity, lasting months to decades, that are interrupted
by short outbursts, during which their X-ray brightness increases by
several orders of magnitude (Bradt et al.\
2000). Figure~(\ref{fig:trans}) shows typical lightcurves of a
neutron-star (Aql~X-1) and a black-hole transient (GRO~J1655$-$40) as
observed by the All Sky Monitor on {\em RXTE} (cf.\ \S4).

The nature of the compact object in a transient system appears to
affect its properties in four ways: the fraction of transients among \mk
the black-hole systems is larger than the fraction of transients among
\mk 
neutron-star systems and their outbursts are typically longer and
rarer \mk(see, e.g., Figure~\ref{fig:trans}); moreover, black-hole
transients in quiescence are significantly fainter than their
neutron-star counterparts. According to our current understanding, the
above differences are caused by the different mass ratios of the
members of the binary systems between the two populations as well as
by the presence of an event horizon in the black-hole systems.

The prevailing model of transient sources is based on the disk
instability model of illuminated accretion disks (van Paradijs 1996;
King, Kolb, \& Burderi 1996): accretion flows that extend to large
radii (typically $>10^{9}-10^{10}$~cm) from the compact object have
characteristic temperatures less than $\sim 10^4$~K, at which the
anomalous opacity related to the ionization of hydrogen renders them
susceptible to a thermal disk instability (see King, \S13). At
the off-cycle of the instability, material piles up at the outer edges
% do we know this? \mk
of the accretion disk with very little, if any, mass accreted by the
central object; this is the quiescent phase of the transient. When the
disk becomes unstable, the accretion flow evolves towards the central
object at the viscous timescale, and the system becomes a bright X-ray
source in outburst.

\begin{figure*}
\centerline{
\psfig{file=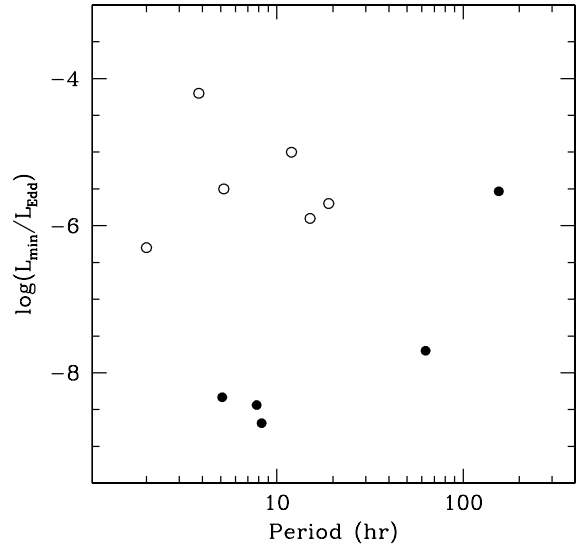,angle=0,width=8truecm}}
 \caption{The quiescent X-ray luminosity of neutron-star (open
circles) and black-hole transients (filled circles) as a function of
their orbital period (after Garcia et al.\ 2001).}
 \label{fig:quiesc}
\end{figure*}

The disk-instability mechanism depends crucially on the temperature of
the accretion disk, and hence heating of the disk by illumination can
alter the above picture. The details of illumination by the central
object or the disk itself are hard to compute, as they depend strongly
on how concave the accretion disk is, on the presence of warps
that can expose different parts of the disk to radiation, and on the
effects of disk winds that can backscatter radiation towards the disk
plane (see, e.g., Dubus et al.\ 1999).  It is generally expected,
though, that neutron stars illuminate their accretion
disks more efficiently than black holes \mk, since most of the accretion energy has to be released near
their surfaces. The stabilizing effect of illumination can then
account for the difference in the outburst properties between
neutron-star and black-hole systems.

Studies of the quiescent emission of X-ray transients became possible
only after the launch of X-ray telescopes with good resolution and low
background, such as {\em ASCA\/} and the {\em Chandra\/} X-ray Observatory
(see, e.g., Rutledge et al.\ 2001). It is now well established that
the quiescent emission of black-hole transients is fainter by more
than an order of magnitude compared to the quiescent emission of
neutron-star sources in similar binary systems
(Fig.~\ref{fig:quiesc}). The time-variability and non-thermal
character of the X-ray brightness in both cases strongly suggests that
at least a large fraction of the emission arises from accretion (see,
e.g., Narayan, Garcia \& McClintock 2001). However, release of heat
buried in the deep layers of a neutron star (Brown, Bildsten, \&
Rutledge 1998) and coronal emission from the companion star (Bildsten
\& Rutledge 2000; Campana \& Stella 2000; but see Lasota 2000) are
also expected to contribute to the total X-ray brightness. In most
cases, the existence of an event horizon in the black-hole sources,
which traps a large fraction of the accretion luminosity and does not
allow the storage of latent heat, is believed to be responsible for
their significantly less luminous quiescent emission (Narayan et al.\
2001).

\begin{table}
 \caption{Masses of Compact Objects$^{a}$}
  \label{tab:masses}
  \begin{tabular}{lcc}
   \hline\hline
   {Source} & {Period} & {Mass}\\
          &  (d)   & ($M_\odot$)\\
   \hline
   Black Hole Candidates & & \\
   \hline
   GRO~J0422$+$32   & 0.212    & 3.66--4.97\\
   A0620$-$00       & 0.323    & 8.70--12.86\\
   GRS~1009$-$45    & 0.285    & 3.64--4.74\\
   XTE~J1118$+$480  & 0.170    & 6.48--7.19\\
   GS~1124$-$683    & 0.433    & 6.47--8.18\\
   4U~1535$-$47     & 1.116    & 8.45--10.39\\
   XTE~J1550$-$564  & 1.543    & 8.36--10.76\\
   GRO~J1655$-$40   & 2.622    & 6.03--6.57\\
   H1705$-$250      & 0.521    & 5.64--8.30\\
   SAX~J1819.3$-$2525 & 2.817  & 6.82--7.42\\
   XTE~J1859$+$226  & 0.382    & 7.6--12.0 (?)\\
   GRS~1915$+$105   & 34       & 10.0--18.0 (?)\\
   GS~2000$+$25     & 0.344    & 7.15--7.78\\
   GS~2023$+$338    & 6.471    & 10.06--13.88\\
   LMC~X-3          & 1.705    & 5.94--9.17\\
   LMC~X-1          & 4.229    & 4.0--10.0 (?)\\
   Cyg~X-1          & 5.600    & 6.85--13.25\\
   \hline
   Accreting Neutron Stars & &  \\
   \hline 
   Vela~X-1         & 8.964    & 1.78$\pm$0.15\\
   4U~1538$-$52     & 3.728    & 1.06$^{+0.41}_{-0.34}$\\
   SMC~X-1          & 3.892    & 1.17$^{+0.36}_{-0.32}$\\
   LMC~X-4          & 1.408    & 1.47$^{+0.44}_{-0.39}$\\
   Cen~X-3          & 2.087    & 1.09$^{+0.57}_{-0.52}$\\
   Her~X-1          & 1.700    & 1.04$^{+0.75}_{-0.58}$ or 1.47$^{+0.23}_{-0.37}$\\
   Cyg~X-2          & 9.844    & 1.78$\pm$0.23\\
   \hline
   \hline
  \end{tabular}\\
 {\footnotesize $^a$Black hole masses are from Orosz (2002); accreting
neutron-star masses are from van Kerkwijk et al.\ (1995), Barziv et al.\
(2001), and Orosz \& Kuulkers (1999); 
radio pulsar masses are from Thorsett \& Chakrabarty (1998).}
\end{table}

\begin{table}
 \caption{Masses of Compact Objects (continued)}
  \begin{tabular}{lcc}
   \hline\hline
   {Source} & {Period} & {Mass}\\
          &  (d)   & ($M_\odot$)\\
   \hline
   Double Neutron Stars & & \\
   \hline
   J1518$+$4904     & 8.634    & 1.56$^{+0.13}_{-0.44}$\\
                    &          & 1.05$^{+0.45}_{-0.11}$\\
   B1534$+$12       & 0.421    & 1.339$\pm$0.003\\
                    &          & 1.339$\pm$0.003\\
   B1913$+$16       & 0.323    & 1.4411$\pm$0.00035\\
                    &          & 1.3874$\pm$0.00035\\
   B2127$+$11C      & 0.335    & 1.349$\pm$0.040\\
                    &          & 1.363$\pm$0.040\\
   B2303$+$46       & 12.34    & 1.30$^{+0.13}_{-0.46}$\\
                    &          & 1.34$^{+0.47}_{-0.13}$\\
   \hline
   Radio Pulsars in Binaries & &  \\
   \hline
   J0437$-$4715     & 5.741    & $<$1.51\\
   J1012$+$537      & 0.605    & 1.7$\pm$0.5\\
   J1045$-$4509     & 4.084    & $<$1.48\\
   J1713$+$0747     & 67.825   & 1.45$\pm$0.31\\
   B1802$-$07       & 2.617    & 1.26$^{+0.08}_{-0.17}$\\
   J1804$-$2718     & 11.129   & $<$1.73\\
   B1855$+$09       & 12.327   & 1.41$\pm$0.20\\
   J2019$+$2425     & 76.512   & $<$1.68\\
   J0045$-$7319     & 51.169   & 1.58$\pm$0.34\\
   \hline\hline
  \end{tabular}
\end{table}
% is pulsar-pulsar system here? \mk

The transient nature of these systems had hampered their systematic
study until recently. Since the mid-1990's, however, the Wide-Field
Cameras on {\em BeppoSAX\/} and the All Sky Monitor on {\em RXTE\/} have
revealed and monitored a large number of transient sources. The impact
of such monitoring programs has been enormous. The statistics of the
transient sources and their recurrence times shed light on \mk the total
number of X-ray binaries in the Galaxy, their birthrates, and
formation mechanisms. The rise and decay times of their outbursts
helped constrain the efficiency of angular momentum transport in
accretion disks (e.g., Hameury et al.\ 1998). More significantly,
however, optical observations of X-ray transients in quiescence
allowed for the measurement of the masses of many compact objects and
have thus provided the best evidence for the existence of stellar-mass
black holes in the Galaxy (McClintock \& Remillard, this volume).

\subsection{Long-wavelength counterparts}

X-ray binaries are initially identified from their intense X-ray
brightness and hard spectra. However, the properties of the binary
systems, such as their orbital periods, the masses of the companion
stars, and the masses of the compact objects can be determined only if a
counterpart of the X-ray source in other wavelengths is also observed
(see Charles, \S5).

The brightness of luminous X-ray sources in optical wavelengths,
especially for compact objects with low-mass companions, is typically
positively correlated with their X-ray brightness and orbital periods
(van Paradijs \& McClintock 1995). This is a strong indication that
thermal emission from the outer parts of the accretion flow as well as
reprocessing of the X-ray emission at the outer accretion disk and the
companion star are responsible for at least part of the optical
emission. 

Detailed studies of X-ray binaries in IR/optical/UV wavelengths are
crucial in measuring the temperature profiles, ionization fractions,
and abundances of elements in the accretion disk (see, e.g., Kallman 
et al.\ 1998) or even the inclinations of the
binary systems (see, e.g., de Jong et al.\ 1996; Wang et al.\ 2001).
Moreover, if the accretion geometry is that of a geometrically thin
disk, such studies most probably provide the only handle in measuring
the rate at which matter is transfered towards the compact object
(see, e.g., Vrtilek et al.\ 1990).

\begin{figure*}[t]
 \centerline{ \psfig{file=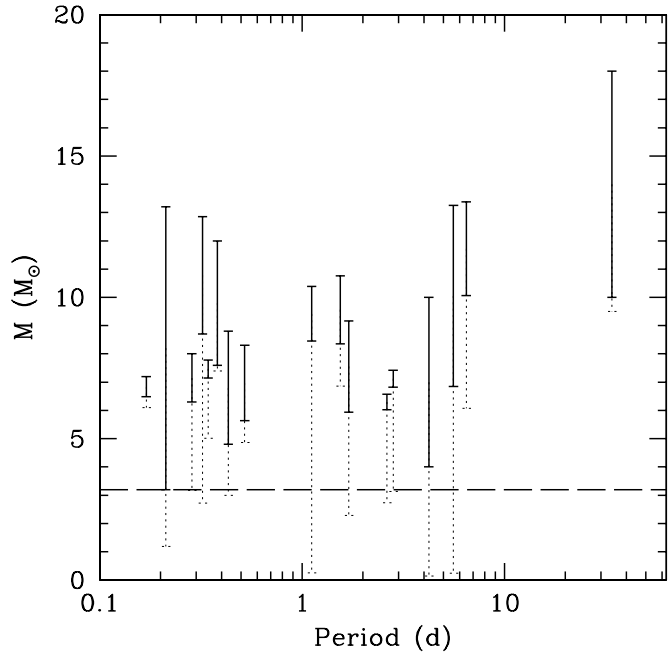,angle=0,width=9truecm}}
 \caption{Masses of black-hole candidates in binary systems and their
 orbital periods; the horizontal dashed line corresponds to the $3.2
 M_\odot$ separatrix (see \S1.3.8) between neutron stars and
 black-hole candidates (after Orosz 2002). }
\label{fig:mass}
\end{figure*}

In transient systems during their quiescent periods as well as in
systems with short orbital periods, the optical emission is dominated
by the companion star, making possible the measurement of their binary
orbits and of the masses of the compact objects. Table~\ref{tab:masses}
and Fig.~\ref{fig:mass} summarize the current mass measurements of
compact objects in accreting binary systems and compare them to the
mass measurements of radio pulsars in non-interacting binaries.
There is a clear dichotomy between compact objects clustered around 
$\sim 1.5 M_\odot$ and objects with significantly larger masses. 
This is believed to be a direct consequence of the fact that there is
an upper limit on the mass of a stable neutron star and any heavier compact
object must be a black hole (but see also \S\ref{sect:nature}).
% but why then dichotomy? \mk

Recent developments in observational methods and analysis techniques
of long-wavelength data have also led to a number of ways of imaging
indirectly the accretion flows around compact objects (see, e.g.,
Harlaftis 2001). Eclipse mapping is applicable only to
high-inclination systems and uses the periodic occultation of
different parts of the accretion flow by the companion star during the
orbit to infer the relative contribution of different regions to the
total emission (see, e.g., Vrielmann 2000 for a review of the method
and applications to cataclysmic variables). Doppler tomography uses the
dependence on orbital phase of Doppler-shifted atomic lines that
originate from different parts of the accretion flow in order to
produce a map of the line-emitting regions on the orbital plane \mk of the
binary system (see, e.g., Marsh 2000; Harlaftis 2001). Finally, echo
tomography uses the time delays between prompt and reprocessed
emission at different regions in order to map the geometry of the
accretion flows and the binary orbits (see, e.g., Horne 2003).

\begin{figure*}
 \centerline{ \psfig{file=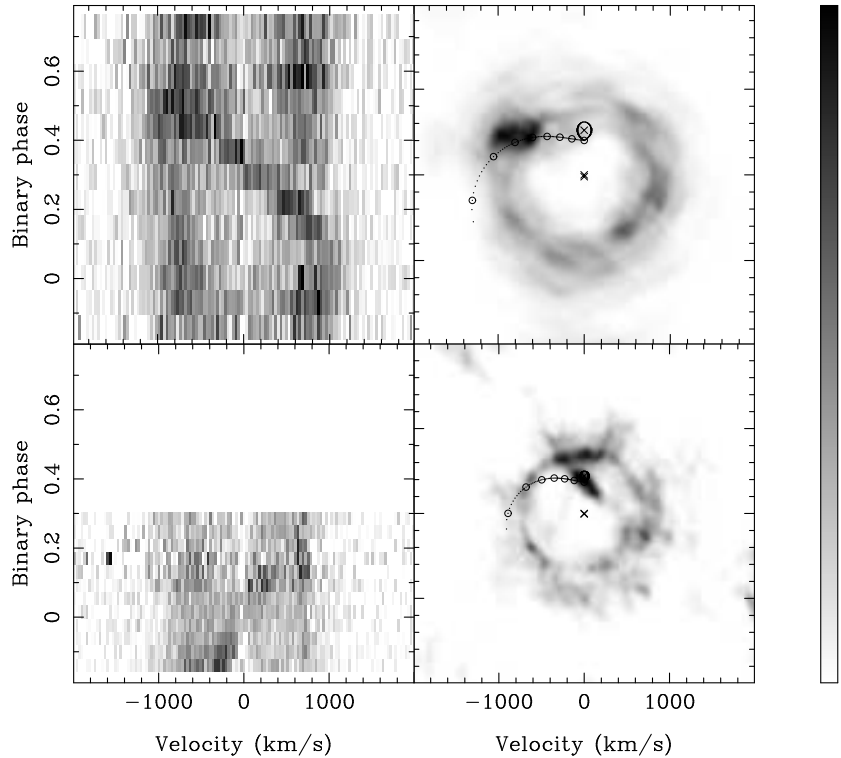,angle=0,width=11truecm}}
 \caption{H$\alpha$ spectra (left panels) and reconstructed Doppler
 images (right panels) of the black-hole candidates GS~2000$+$25 (top)
 and Nova Oph~1977 (bottom; Harlaftis 2001).}
 \label{fig:doppler}
\end{figure*}

The above indirect imaging techniques provide the most concrete method
of observationally testing accretion flow models. They have been
regularly used in mapping the radial temperatures of geometrically
thin accretion disks, often showing significantly flatter profiles
compared to accretion theory (see Vrielmann 2000). They can be used to
measure the geometrical thickness of the accretion flows, thereby
distinguishing between competing models (see, e.g., O'Brien et al.\
2002).  Finally, they provide the best evidence for the presence of
spiral structures in accretion disks, thus giving clues to the viscous
mechanisms that operate in these flows (see, e.g., Harlaftis 2001).

\subsection{Jets}

\begin{figure*}
 \centerline{\psfig{file=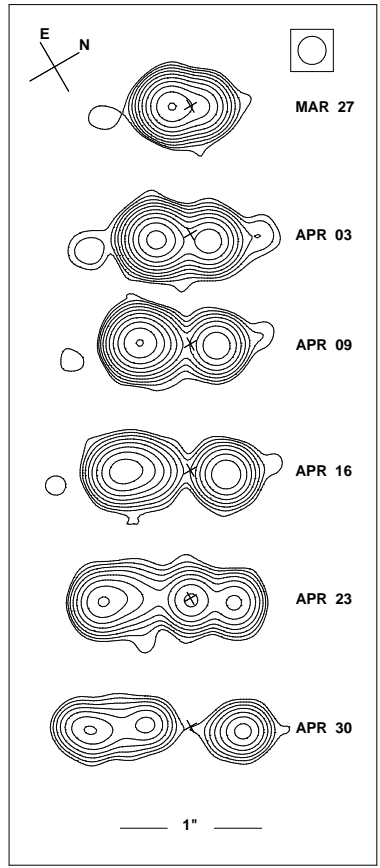,angle=0,width=5.truecm}} \caption{A
 series of 3.5cm VLA images of the black-hole candidate source
 GRS~1915$+$105, showing the fast-traveling knots of the radio jet 
 (Mirabel \& Rodriguez 1994).} 
\label{fig:jet}
\end{figure*}

Outflows and collimated jets are ubiquitous phenomena in all accreting
objects, from young stars to active galactic nuclei.  The discovery of
mildly relativistic jets from the binary system SS~433 (Spencer 1979)
and of superluminal jets from the black-hole candidate GRS~1915$+$105
(Mirabel \& Rodriguez 1994; also Fig.~\ref{fig:jet}) revealed that
accreting galactic compact objects are no exception to this rule (see
Fender, \S9).

Jets from accreting stellar-mass black holes in the galaxy share many
properties with their counterparts in active galactic nuclei. For
example, they have non-thermal, polarized radio spectra indicating the
presence of shock-accelerated relativistic electrons that emit
synchrotron radiation as they propagate in regions with large-scale
magnetic fields (e.g., Mirabel et al.  1998; Eikenberry et al.\
1998). They also show large flux ratios between the approaching and
receding sides of the jets, as expected for relativistic flows
(Mirabel \& Rodriguez 1994).

In addition to revealing such similarities, however, the short
dynamical timescales associated with galactic black-hole jets and
their proximity made possible an in depth study of their properties,
even though only a handful of jet sources is currently known. The
orientation of some jets in the sky, with the jet in SS~433 being the
main example (Margon 1984), has been observed to precess in real time
over large angles with long periods ($\sim 162$ days in the case of
SS~433). This is believed to be associated to the precession of the
underlying accretion disk (see, e.g., Begelman \& Rees 1984; Ogilvie
\& Dubus 2001) and may provide clues towards understanding the launching
and collimation of the jets.

The propagation speeds of the jets and their deceleration upon
interactions with the interstellar medium can be measured on images by
following the kinematics of individual radio knots (see, e.g,
Fig.~\ref{fig:jet}).  Not surprisingly, the inferred speeds appear to
depend on the state of the underlying accretion disk from which they
are launched, with steady and transient sources showing mildly and
highly relativistic jets, respectively (see also Fender, \S9\mk).

Coordinated observations of accreting black holes and neutron stars in
the radio, infrared, and X-rays recently revealed \mk the most striking of
jet properties: there is a very clear correlation between the presence
of jets and the X-ray spectral state of the accretion flows (see,
e.g., Corbel et al.\ 2000). In particular, jets appear when the X-ray
spectra of the sources indicate emission from hot electrons ($\sim
100$~keV; see next section); on the other hand, when their spectra are
typical of cold, geometrically thin accretion disks, the jets are weak
or absent.  It is unclear at this point what is the causal connection
between the radio and X-ray properties of accreting compact
objects. The mechanism responsible for the heating of electrons in the
accretion flow may be related to the formation of an outflow, as
is the case both for magnetically active accretion disks (see, e.g.,
Blandford \& Payne 1982) or for advection dominated accretion flows
(Narayan \& Yi 1994). Alternatively, most \mk of the X-ray
emission may be produced directly at the base of the jet (Markoff et
al.\ 2001).

Finally, recent observations of the X-ray emission from the jet of
SS~433 with {\em Chandra\/} confirmed the presence of atomic emission
lines, strongly suggesting that heavy ions are also accelerated
together with the electrons and positrons that are responsible for
most of the jet emission (Marshall et al.\ 2002).  The jet in SS~433,
however, is only mildly relativistic and it is not clear whether this
is related to the presence of ions in the jet or if there simply is
\mk an observational selection effect against detecting lines from
very relativistic outflows (Mirabel et al.\ 1997).

\begin{figure*}
 \centerline{\psfig{file=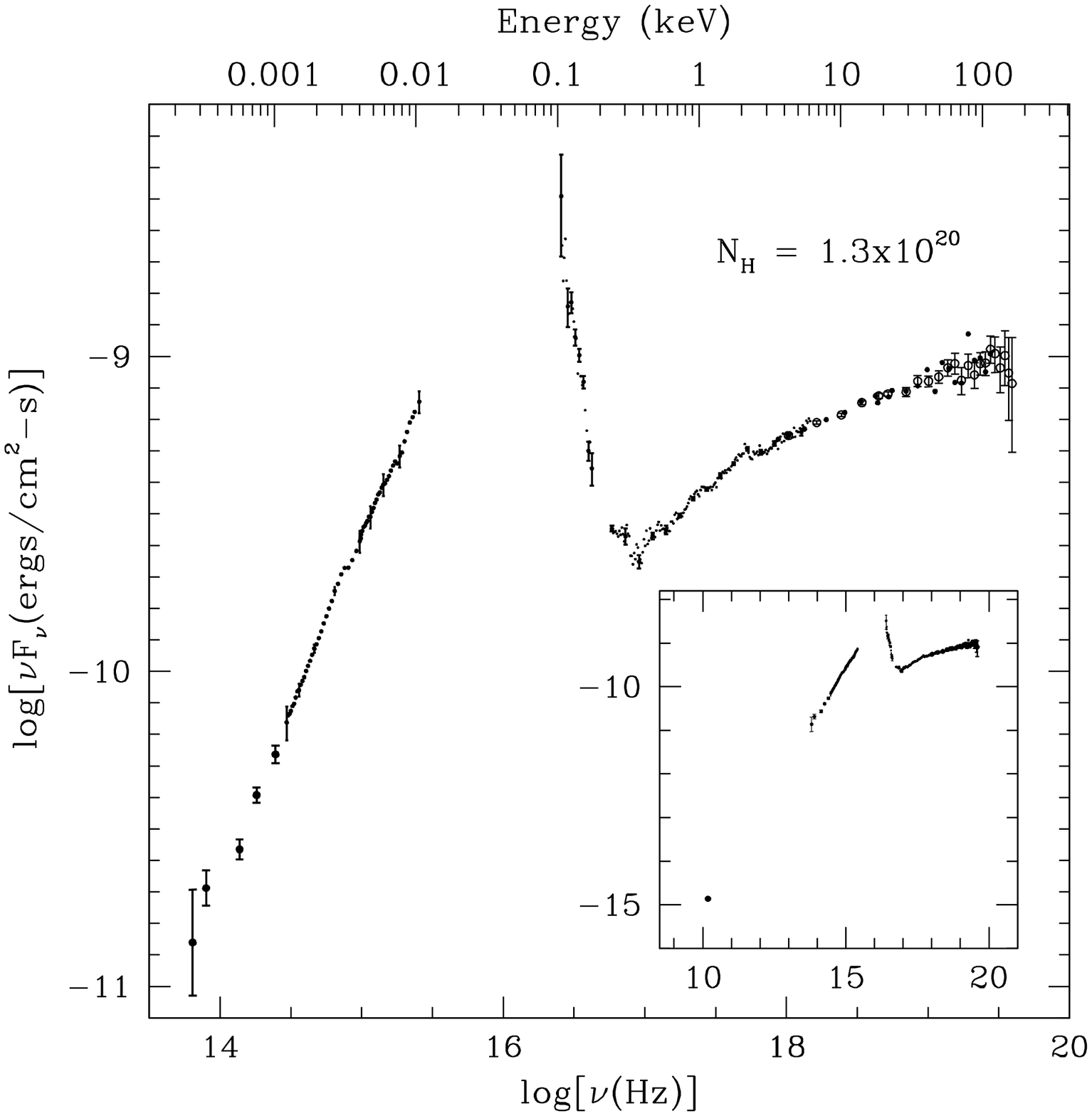,angle=0,width=6.8truecm}
 \psfig{file=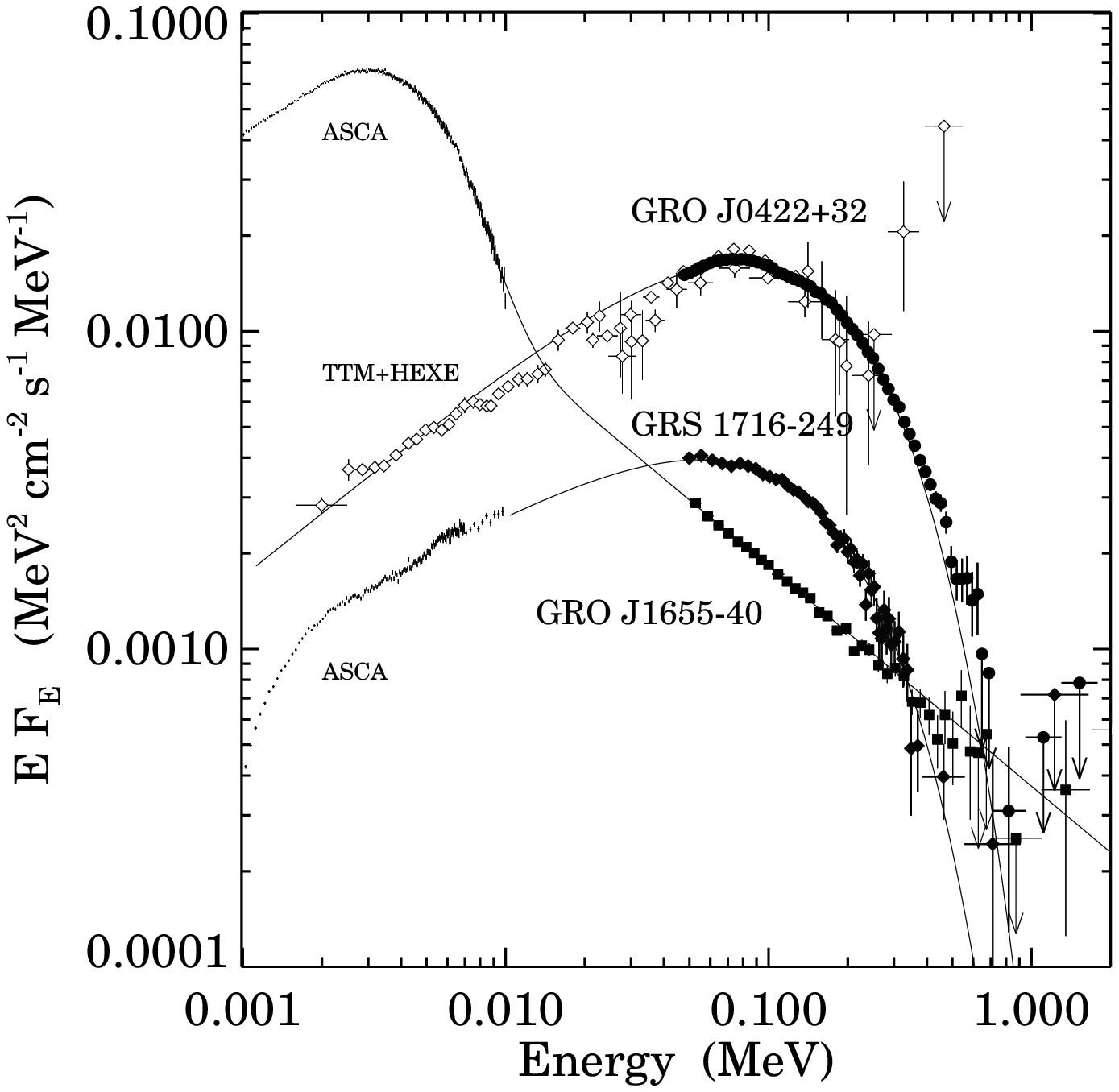,angle=0,width=6.8truecm}} 
\caption{Broad-band spectra of black-hole candidates: (left) The 
infrared to soft X-ray spectrum of XTE~J1118$+$480 (McClintock et al.\
2001); (right) the soft X-ray to $\gamma$-ray spectra of three sources
(Grove et al.\ 1998).}
\label{fig:broadsp}
\end{figure*}

\subsection{X-ray and $\gamma$-ray spectroscopy}

The design of high-energy missions with broad-band spectral coverage,
the numerous campaigns of simultaneous observations with multiple
instruments, as well as the advent of high-resolution CCD and grating
spectrographs for X-rays have launched a new era of high-energy
spectroscopy in astrophysics. Accreting compact objects are often
being monitored in all wavelengths, providing a strong handle on their
bolometric luminosities as well as placing stringent constraints on
accretion models.

Figure~\ref{fig:broadsp} shows some examples of broad-band spectra
of black-hole candidates, from the infrared to the $\gamma$-rays.  All
such spectra of accreting compact objects show unequivocally that a
number of distinct emission mechanisms are responsible for their
various features. In fact, simple thermal emission models from
geometrically thin accretion disks can produce neither the spectral
complexity nor the hard X-ray and $\gamma$-ray fluxes that are
observed. Modeling \mk the spectra of accreting compact
objects is complicated and appears to be mostly data driven. However,
it is also potentially very rewarding as it may lead to the understanding
of processes such as the generation of magnetic fields \mk in turbulent
flows, viscous heating in magnetic media (see, e.g., Quataert \& Gruzinov
1999), and the thermal properties of multi-temperature plasmas (e.g.,
Coppi 1999).

Most current models of the X-ray emission from accreting neutron stars
and black holes (see GRO~J0422$+$32 and GRS~1716$-$249 in
Fig.~\ref{fig:broadsp}) require that a tenuous atmosphere of hot
($\sim 10-100$~keV) electrons is present simultaneously with the
geometrically thin accretion disks.  Such a hot medium may be 
in the form of a magnetically heated corona (see, e.g., Dove, Wilms,
\& Begelman 1997), of an advection-dominated accretion flow (ADAFs; Esin,
McClintock, \& Narayan 1997), or even a jet (Markoff, Falcke \& Fender 2001). These same electrons are almost certainly responsible
also for the longer-wavelength emission (see XTE~J1118$+$480 in
Fig.~\ref{fig:broadsp}), which appears to be strongly correlated with
the X-ray flux, via radiation processes associated with the magnetic
fields generated and sustained by the accretion flow. Finally, the
power-law $\gamma$-ray spectral tails of luminous neutron-star and
black-hole candidates (see GRO~J1655$-$40 in Fig.~\ref{fig:broadsp};
see also Grove et al.\ 1998; Di Salvo et al.\ 1999) require the
existence of a non-thermal population of very energetic electrons,
either in the form of a hybrid plasma (Coppi 1999) or of a
quasi-radial high-velocity flow (Laurent \& Titarchuk 1999).

\begin{figure*}
\centerline{\psfig{file=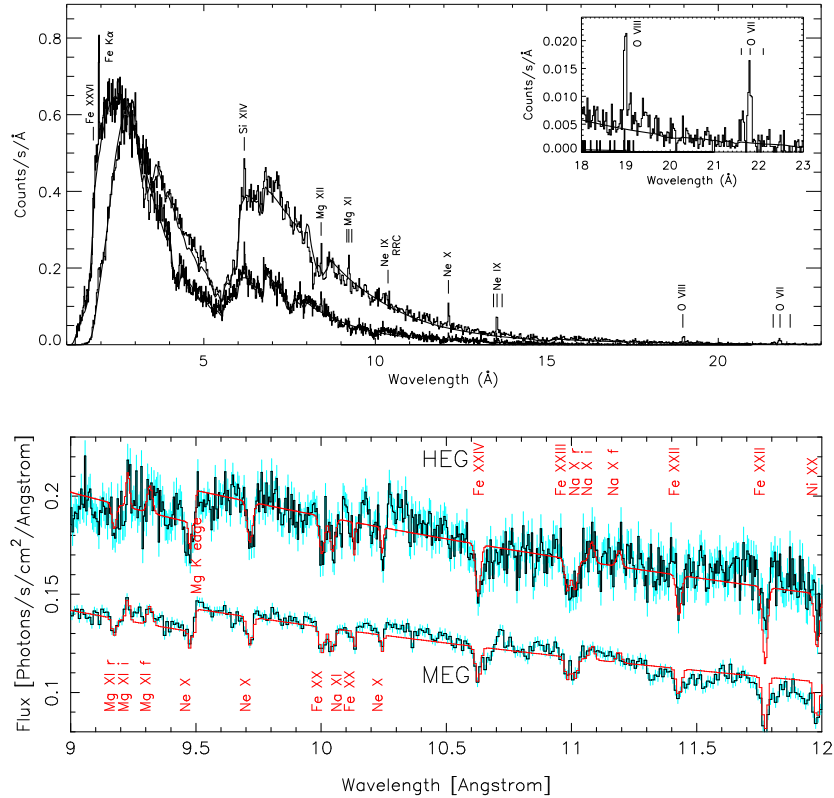,angle=90,width=11.8truecm}}

\mbox{}

\centerline{\psfig{file=cygsp.ps,angle=0,width=11truecm}}
 \caption{High resolution X-ray spectra of {\em (top)\/} the ADC
 source 4U~1822$-$37 (Cottam et al.\ 2001) and ({\em bottom\/}) the
 black-hole source Cyg~X-1 as observed by the {\em Chandra\/} X-ray
 Observatory (Miller et al.\ 2002).}  \label{fig:linesp}
\end{figure*}

Perhaps the most eagerly \mk anticipated result of the launch of X-ray telescopes
with high spectra resolution, such as {\em ASCA}, {\em Chandra}, and
{\em XMM/Newton}, has been the discovery of atomic lines from the
spectra of accreting compact objects. The relative strengths and
equivalent widths of such lines depend strongly on temperature,
density, and ionization flux and hence are valuable probes of the
physical conditions in the accretion flows (Liedahl 1999). Moreover,
the gravitational redshifts and relativistic broadening of atomic
lines generated close to the event horizons of black holes can, in
principle, be used to map the spacetimes around the compact objects
and measure properties such as their masses and spin angular momenta
(e.g., Fabian et al.\ 1989).

The X-ray spectra of many accreting compact objects have,
unfortunately, atomic lines that are very weak or even undetectable\mk, largely due to their
high temperatures and photonization fluxes. However, studies of atomic
lines have been proven fruitful in several cases for which the binary
configurations are optimal. For example, in accretion-disk corona (ADC)
sources, the high inclinations of the binary systems allow for a clear
view of the coronal structure away from the central object, where the
temperatures are lower and the line emission stronger (see
Fig.~\ref{fig:linesp}; also Cottam et al.\ 2001). In binary systems
with companions that exhibit strong winds, the X-ray spectra show a
variety of absorption lines and edges that originate at the relatively
cooler wind material (Fig.~\ref{fig:linesp} and, e.g., Miller et al.\
2002a). Recent observations of several black-hole candidates have also
shown evidence for relativistically broadened iron K lines, similar to
those observed in active galactic nuclei (Miller et al.\ 2002b).
Finally, a detection of gravitationally redshifted atomic lines during
thermonuclear flashes on the surface of an accreting \mk neutron star has
also been recently reported (Cottam et al.\ 2002).

\subsection{Variability}

\begin{figure*}
 \centerline{ \psfig{file=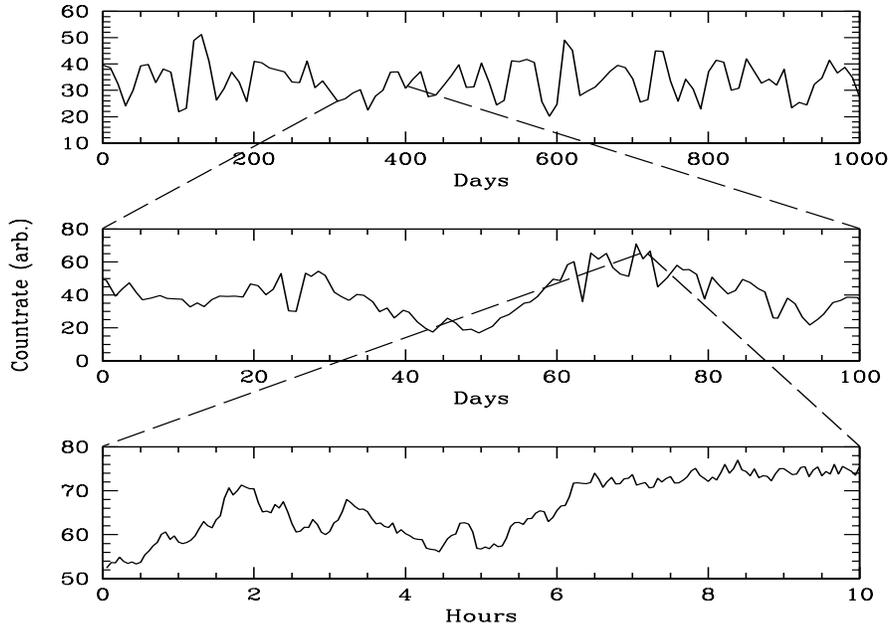,angle=0,width=12truecm,height=9truecm}}
 \caption{The light curve of the neutron-star source Cyg~X-2, as observed 
 at different timescales by the All Sky Monitor (upper two panels) and the 
  Proportional Counter Array (bottom panel) onboard {\em RXTE\/}.}
 \label{fig:long}
\end{figure*}

Accreting compact objects are among the most variable persistent
sources in the sky. Even excluding the X-ray transients, the flux from
most sources typically varies by factors of two over periods ranging
from months to fractions of a second (see Fig.~\ref{fig:long}). This
is not surprising, given the wide range of characteristic timescales
that are involved in the processes that lead to the production of the
high-energy emission. For example, the transfer of mass from the
binary companion to the compact object is expected to vary at
timescales comparable to the orbital period, i.e., hours to days. The
inward diffusion of matter in the accretion flow occurs at the viscous
timescale, which is slower than the Keplerian orbital frequency at any
radius and \mk ranges from days, at the outer edge of the disk, to
fractions of a second, close to the compact object. Finally, the
interaction of the accretion flow with the central star occurs at the
fastest dynamical timescale in the accretion flow, which is of order
of a millisecond.

\begin{figure*}
 \centerline{ \psfig{file=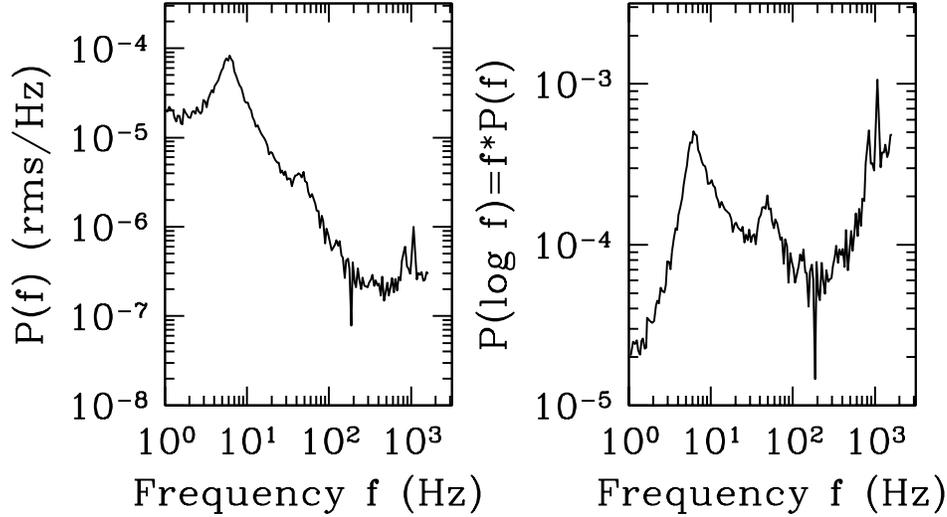,angle=0,width=14truecm}}
 \caption{A power-density spectrum of the neutron-star source Sco~X-1
 (in two commonly used representations) showing multiple,
 simultaneously detected quasi-periodic oscillations.}
 \label{fig:qpo}
\end{figure*}

Over the last two decades, the most unexpected result of timing
studies of accreting compact objects has been the discovery of
quasi-periodic oscillations (QPO; see Fig.~\ref{fig:qpo}) of their
X-ray brightness at all these timescales (van der Klis 2000, and
\S2). Because of the high degree of variability of the sources as well
as due to observational constraints, the fast (greater than a fraction
of a Hz) oscillations of the X-ray brightness are the ones that have
been studied more extensively, mostly with the proportional counters
onboard {\em EXOSAT}, {\em GINGA}, and {\em RXTE}.

The large majority of these oscillations have frequencies that are
highly variable, even though they may loose coherence only after tens
or hundreds of cycles. When several \mk variable-frequency QPOs are
observed simultaneously, their frequencies follow a small number of
tight correlations (Psaltis, Belloni, \& van der Klis 1998). On the
other hand, in several luminous black-hole candidates, pairs of
high-frequency QPOs have also been detected, with frequencies that are
nearly constant and in small integer ratios (i.e., 3:2, 5:3;
Strohmayer 2001a, 2001b; Abramowicz \& Kluzniak 2001). These distinct
and correlated frequencies of the observed QPOs clearly suggest that
accretion flows are capable of picking, out of a large pool of
alternatives, only a small number of characteristic frequencies at
which to vary preferentially.

A number of theoretical interpretations have been put forward so far,
in an attempt to account for the mechanism that picks these
characteristic frequencies in the accretion flows. Current models of
the {\em variable-frequency\/} QPOs require the presence of a
characteristic radius, across which the properties of the accretion
disks change considerably. The observed QPO frequencies are then
attributed to the characteristic dynamic or hydrodynamic frequencies
of the accretion flow at that radius (e.g., Stella, Vietri \& Morsink
1999; Titarchuk et al.\ 1999; Psaltis \& Norman 1999) or, in the case
of neutron stars, to the coupling of these frequencies to the stellar
spin (Miller, Lamb \& Psaltis 1998). On the other hand, models of
the {\em constant-frequency\/} QPOs in black-hole systems are based on
the trapping of oscillatory modes in accretion disks caused by the
properties of the relativistic spacetime (Wagoner 1999; Kato 2001) or
on the non-linear coupling of different frequencies near the
black-hole event horizons (Abramowicz \& Kluzniak 2001).
% or just motion at the isco \mk

A common feature of all current models of QPOs is the identification
of at least some of the observed frequencies with dynamic frequencies
in the accretion flows, such as the ones related to the azimuthal
orbital motion of plasma, to Lense-Thirring precession, etc. These
frequencies correspond to regions very close to the neutron-star
surface or the black-hole event horizons and hence offer the
possibility of observing, for the first time, effects that occur only in
strong gravitational fields.  In particular, the general relativistic
prediction of a radius, inside which no stable circular orbits exist
(the so-called innermost stable circular orbit or ISCO), is of
fundamental importance for almost all models. This characteristic
radius is responsible, in different models, for the saturation of the
observed QPO frequencies with accretion rate (e.g., Miller et al.\
1998; Zhang et al.\ 1998) or the trapping of modes in the inner
accretion flows and hence the generation of the QPOs themselves
(Wagoner 1999; Kato 2001).

\begin{figure*}
 \centerline{ \psfig{file=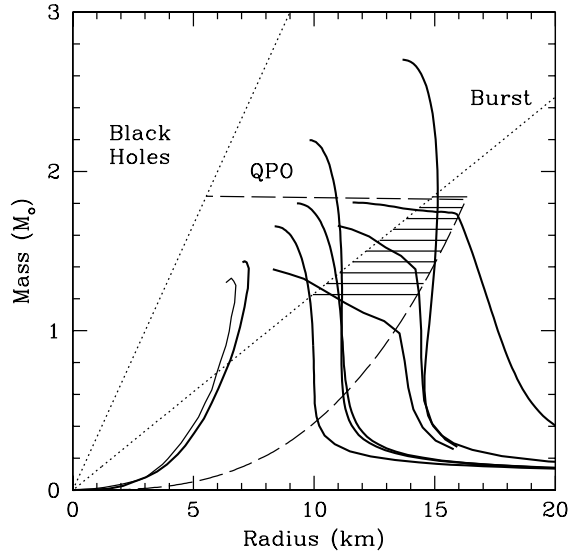,angle=0,width=8truecm}}
 \caption{Constraints on the mass and radius of the neutron star in
 4U~1636$-$36 imposed by the observation of a 1193~Hz QPO (dashed
 line; Jonker, M\'endez \& van der Klis 2002; Miller et al.\ 1998) and
 the large amplitudes of burst oscillations (dotted line; Nath,
 Strohmayer \& Swank 2002). The hatched region represents the area of
 the parameter space that is consistent with all observations and
 includes stars with baryonic masses larger than $\simeq 1.4
 M_\odot$. The solid lines correspond to neutron-star models with
 different equations of state. Stellar models with condensates
 correspond to mass-radius relations with a characteristic flattening
 at small radii; models of strange stars correspond to mass-radius
 relations that start at the origin.}  \label{fig:eos}
\end{figure*}

More importantly, the azimuthal orbital frequency at the radius of the
innermost circular orbit is thought to provide a natural upper limit
on the frequency of any dynamical process that can occur in an
accretion flow (see, e.g., Kluzniak, Michelson \& Wagoner 1990;
Miller et al.\ 1998). Since this azimuthal frequency depends only on
the mass and spin of the compact object, it provides, when compared to
an observed QPO frequency, an upper limit on the mass (modulo the
spin) of the neutron star or black hole. This is illustrated as a
horizontal dashed line in Figure~\ref{fig:eos}, for the case of the
neutron-star source 4U~1636$-$36. An additional upper limit on the
radius of the neutron-star (as a function of its mass) is imposed by
the requirement that the maximum observed QPO frequency is less than
the azimuthal orbital frequency at the stellar surface (see
Fig.~\ref{fig:eos}). These two arguments, together with additional
bounds imposed by the presence of oscillations during thermonuclear
bursts (Nath et al.\ 2002), the detection of redshifted lines from the
stellar surfaces (Cottam et al.\ 2002), and the measurement of the mass
of the compact objects in the binaries using orbital dynamics will be
able to provide the most stringent constraints on the properties of
neutron-star matter and its equation of state.

As discussed above, the rapid variability properties of accretion
flows provide useful probes into the physical conditions close to the
compact objects. At the same time, the slow ($\ge 1$~day) variability
of the same systems can be used in constraining models of the
accretion flows at large distances and of the mass transfer process
between the members of the binary. Such studies have become possible
recently with the systematic observation of the entire X-ray sky using
the All Sky Monitor onboard {\em RXTE\/}.

The long-term variability of accreting weakly-magnetic neutron stars
and black holes is typically aperiodic and reflects the variable
nature of the mass-transfer and accretion processes. Sources with more
systematic slow variability fall in three categories: (a) systems with
large amplitude variations of their X-ray flux because of their
transient nature (see Fig.~\ref{fig:trans}), (b) systems with a
periodic modulation at their orbital frequency caused e.g., by
eclipses, by the reflection of the X-ray photons off the binary
companion and the accretion stream, or by the variable rate of mass
transfer due to a highly elliptical orbit (e.g., Cir~X-1: Shirey et
al.\ 1996), and (c) systems with quasi-periodic modulations at
variable, superorbital periods (see, e.g., Wijnands, Kuulkers \& Smale
1996; Heinz \& Nowak 2001).

\begin{table}
 \caption{Long-term periods of X-ray binaries}
  \label{tab:periods}
  \begin{tabular}{lccl}
   \hline\hline
   {Source} & {Orbital Period} & {Long-term Period} & References\\
          &  (d)   & (d) &\\
   \hline
   \hline
   Stable & & &\\
   \hline
   LMC~X-4 & 1.4 & 30.4 & \\
   Her~X-1 & 1.7 & 35   & \\
   SS 433  & --- & 164$^a$  & 16\\
   \hline
   Quasi-periodic & & &\\
   \hline
   SMC~X-1       & 3.89  & 50--60         & 9 \\
   LMC~X-3       & 1.7   & $\sim$100--300 & 11, 12\\
   Cyg~X-1       & 5.6   & 294            & 13\\
   Cyg~X-2       & 9.84  & $\sim$70--80   & 1, 2, 3, 14, 15\\
   Cen~X-3       & 2.09  & $\sim$120      & 7\\
   4U~1728$-$34  & ---   & $\sim$30--70   & 3, 4\\
   4U~1820$-$30  & 0.008 & $\sim$170      & 6, 10\\ 
   4U~1916$-$053 & 0.035 & $\sim$80       & 5\\
   4U~2127$+$119 & 0.71  & $\sim$37 (?)   & 8\\
   \hline\hline
  \end{tabular}\\
  {\footnotesize $^a$as inferred from the precession of the
jet;\\
  References: 
  1. Kuulkers et al.\ 1996; 
  2. Paul et al.\ 2000; 
  3. Kong et al.\ 1998; 
  4. Galloway et al.\ 2003; 
  5. Homer et al.\ 2001; 
  6. Chou \& Grindlay 2001;
  7. Priedhorsky \& Terrell 1983;
  8. Corbet et al.\ 1997;
  9. Wojdowski et al.\ 1998
 10. Priedhorsky \& Terrell 1984;
 11. Cowley et al.\ 1991;
 12. Wilms et al.\ 2001;
 13. Priedhorsky et al.\ 1983;
 14. Smale \& Lochner 1992;
 15. Kuulkers et al.\ 1999; 
 16. Margon \& Anderson 1989.
  }
\end{table}
\mbox{}

\begin{figure*}
 \centerline{ \psfig{file=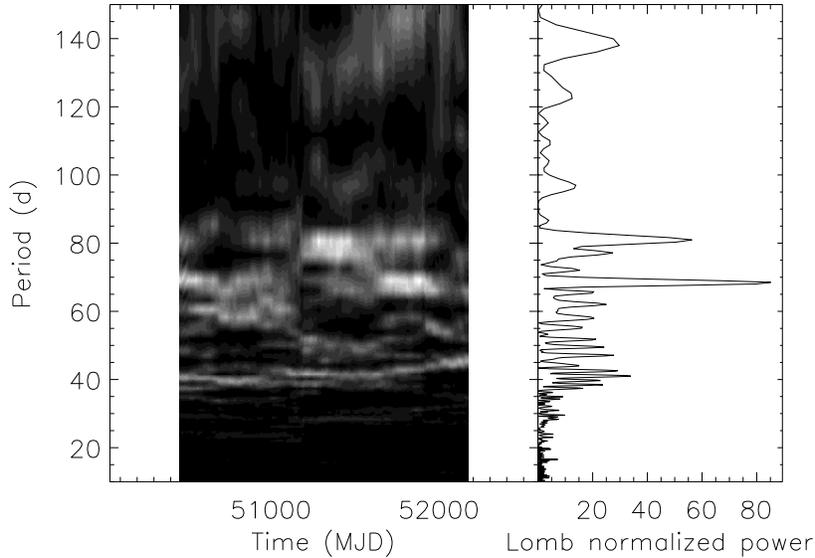,angle=0,width=11truecm}}
 \caption{Dynamical periodogram of the neutron-star source Cyg~X-2,
 showing the variable, superorbital oscillations of its X-ray flux, as
 observed by the All Sky Monitor on {\em RXTE\/} (courtesy D.\ Galloway).}
 \label{fig:warp}
\end{figure*}

Figure~\ref{fig:warp} shows an example of a variable period modulation
in the source Cyg~X-2. This is a neutron-star system \mk with a binary
period of 9.84~days, which shows a number of superobital periods, some
of which are nearly integer multiples of 9.84~days (see also Wijnands
et al. 1996). Such modulation may be related to variable mass transfer
of the binary companion that is driven at the binary orbital period or
to reflection off the warped surface of the outer accretion
disk. Recent theoretical investigations of the warping of
geometrically thin accretion disks caused by the torque of the
reflecting X-ray irradiation from the central object or by an asymmetric
wind (see, e.g., Pringle 1996; Malloney, Begelman \& Pringle 1998)
have indeed shown the possibility of long-lived warping modes that
could produce, in principle, the observed modulations.

\subsection{Thermonuclear Bursts}

\begin{figure*}
 \centerline{ \psfig{file=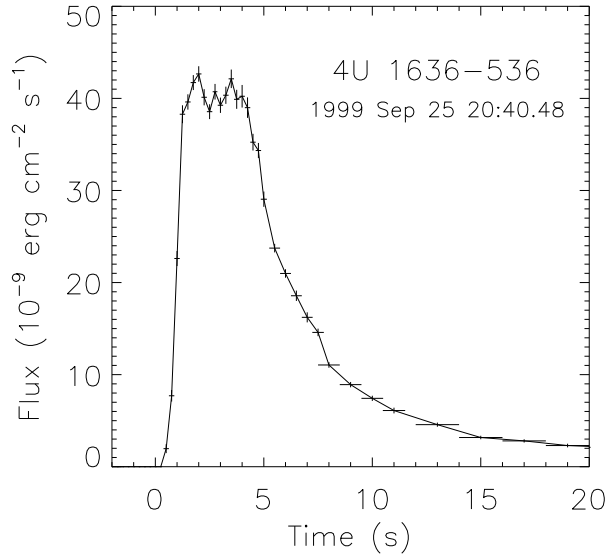,angle=0,width=8truecm}} \caption{A
 typical Type~I X-ray burst from the neutron-star source
 4U~1636$-$36. The flat top of the burst lightcurve is characteristic
 of Eddington-limited (or radius-expansion) bursts (courtesy D.\
 Galloway).}  \label{fig:burst}
\end{figure*}

The material that is accreted on the surface of a weakly-magnetic
neutron star may be compressed to densities and temperatures for which
the thermonuclear burning of helium is unstable. The ignition of
helium results in a rapid ($\sim 1$~s) increase in the X-ray
luminosity of the neutron star, followed by a slower ($\sim$tens of
seconds) decay that reflects the cooling of the surface layers that
ignited (see, Fig.~\ref{fig:burst}). The observational manifestation
of these thermonuclear flashes are called Type~I X-ray bursts (for a
review see Lewin et al.\ 1996 and Strohmayer \& Bildsten, \S3).

For very energetic bursts, the force of the escaping radiation
balances gravity, causing the surface layers of the neutron star to
expand rapidly. During the expansion phase, the emerging radiation
flux remains comparable to the Eddington critical value, at which
radiation and gravitational forces are balanced, and the remaining
energy of the explosion is given as kinetic and potential energy to
the expanding layers (Kato 1983; Nobili et al.\ 1994). These are the
so-called Eddington-limited or photospheric radius-expansion
bursts. Because the maximum observed flux of an Eddington-limited
burst depends, to zeroth order, on the mass and radius of the star
(the allowed values of which span a very narrow range) as well as on
the distance to the source, these burst can be very useful in
constraining all these three parameters and, in particular, in
measuring the distances to X-ray bursters (see Kuulkers et al.\ 2003;
Galloway et al.\ 2003).

\begin{figure*}
 \centerline{ \psfig{file=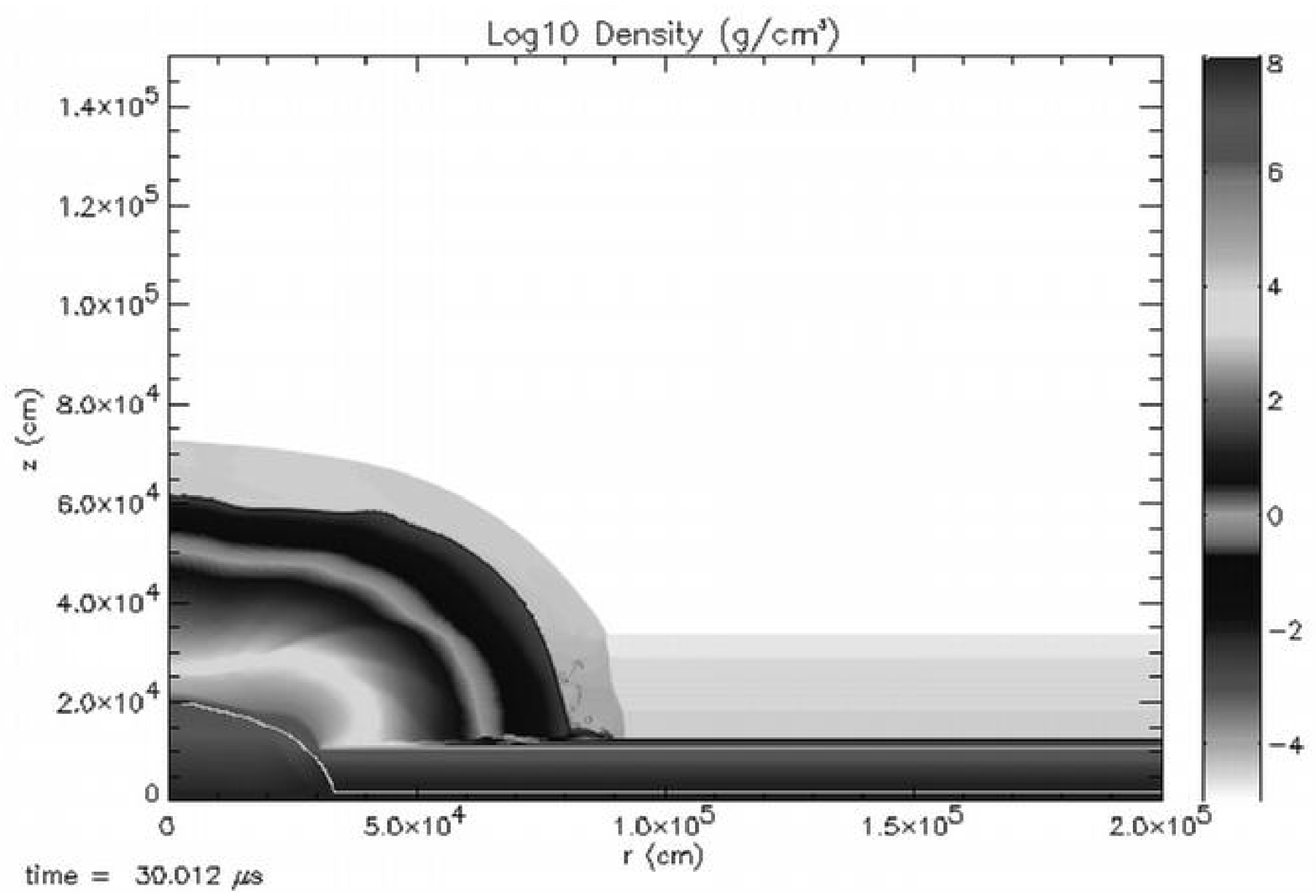,angle=0,width=6.5truecm}
 \psfig{file=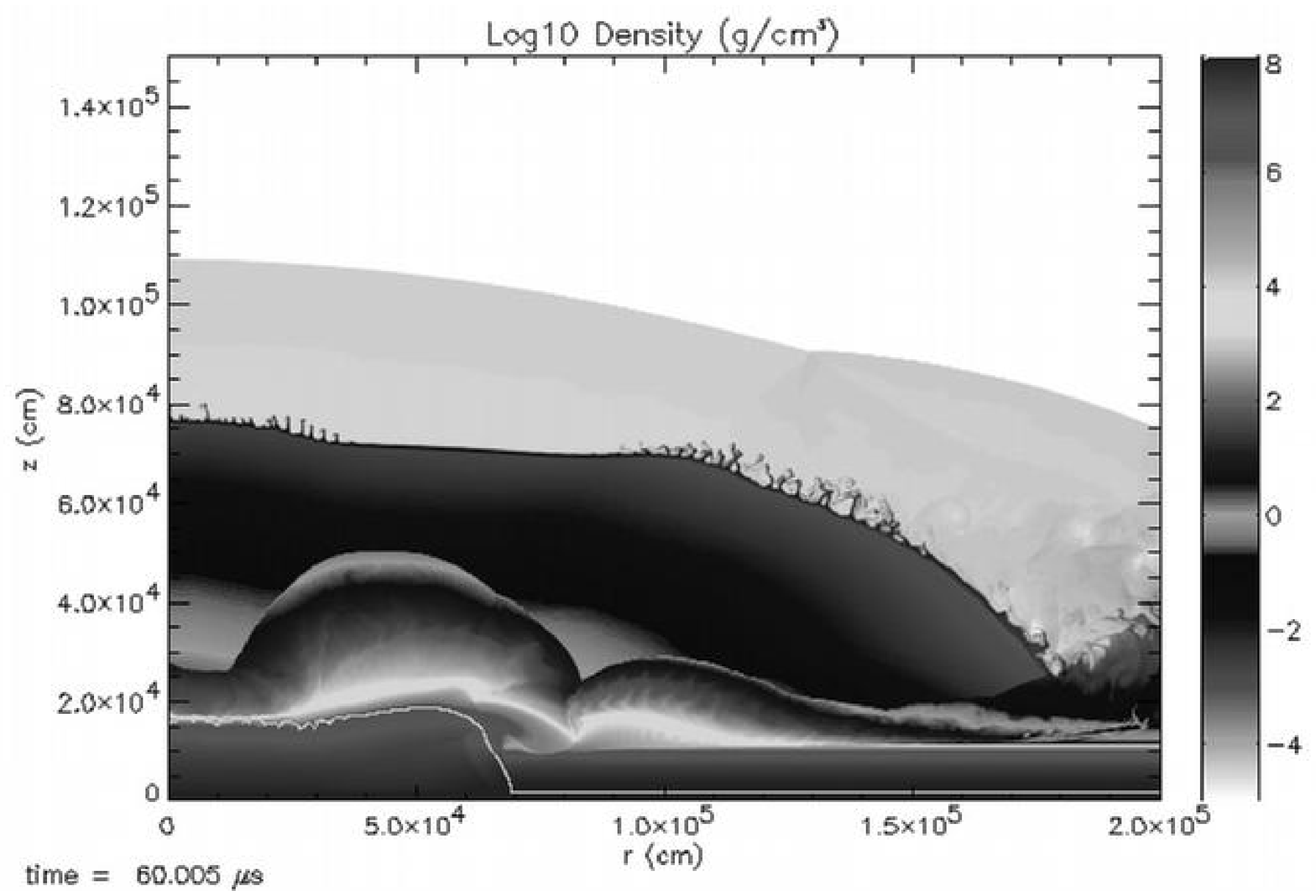,angle=0,width=6.5truecm}} \centerline{
 \psfig{file=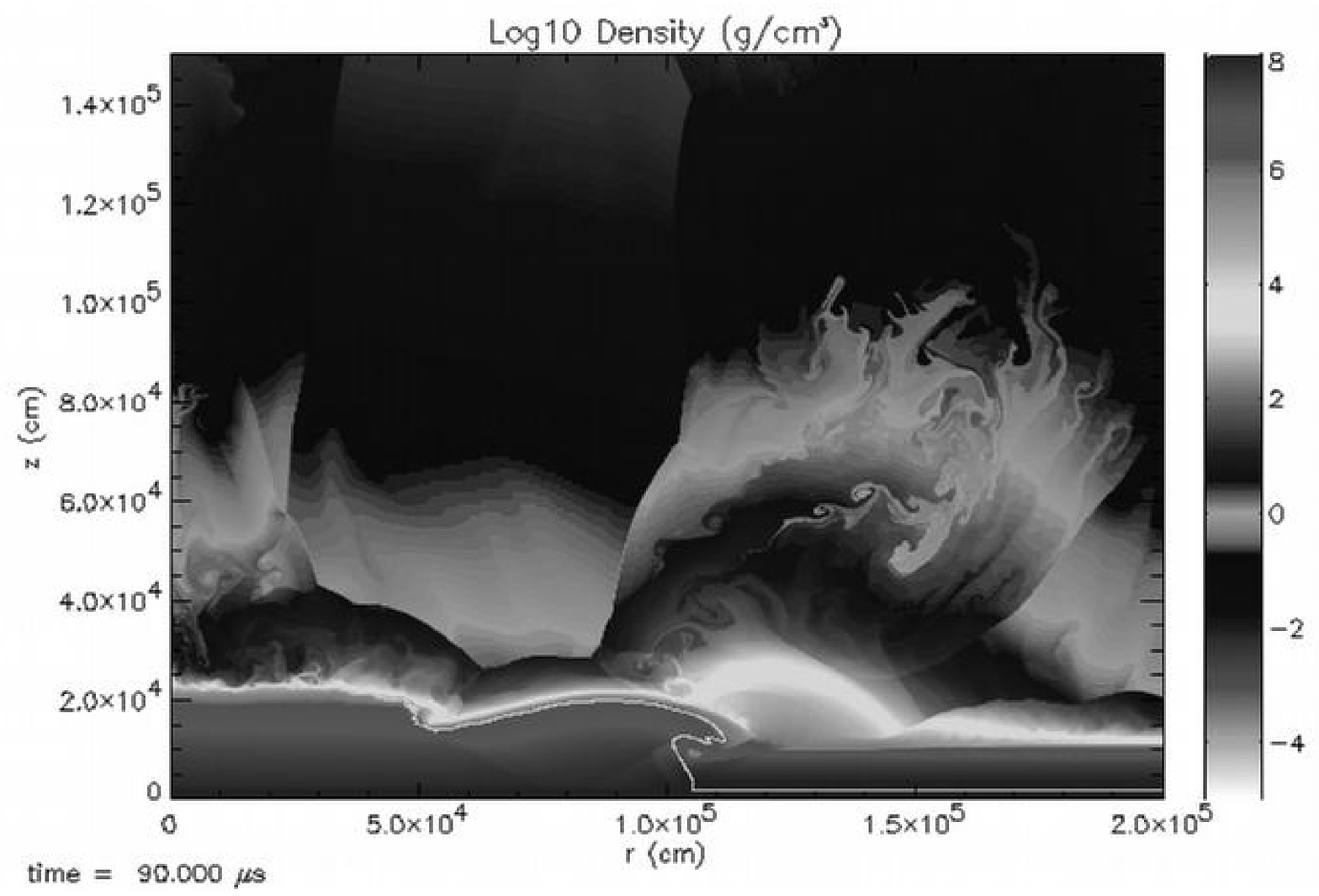,angle=0,width=6.5truecm}
 \psfig{file=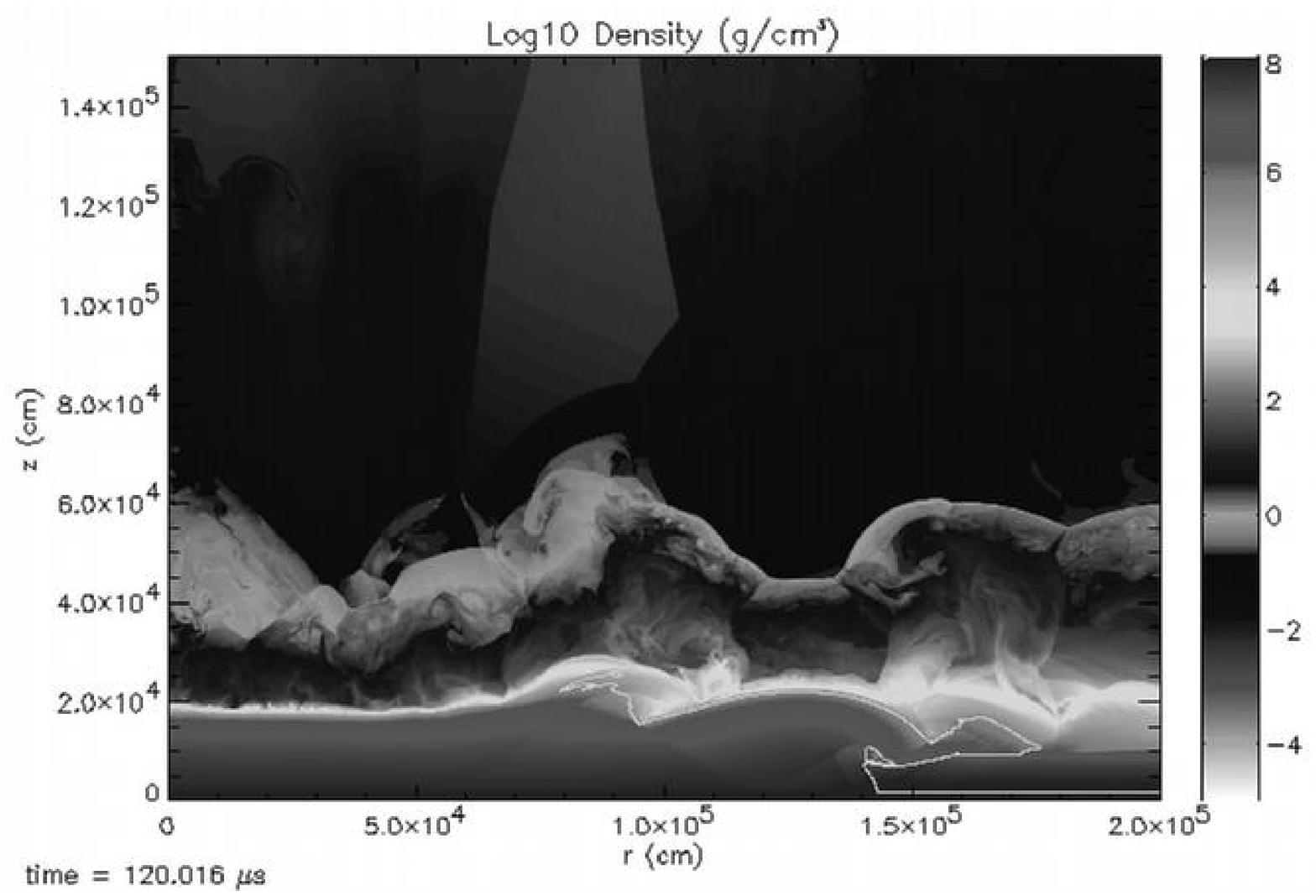,angle=0,width=6.5truecm}}
 \caption{Two-dimensional simulation of helium detonation on a neutron
 star. The greyscale levels correspond to different densities
 (Zingale et al.\ 2001)} \label{fig:burn}
\end{figure*}

The general properties of Type~I X-ray bursts, such as their
energetics, peak fluxes and fluences, recurrence times, rise and decay
timescales, etc.\ are qualitatively consistent with the predictions of
the helium ignition model. Quantitatively, however, the observations
show very little of the regularity that is inherent in the numerical
models (Lewin et al.\ 1996; see, however, Galloway et al.\ 2004). This
is not surprising given the strong dependence of the burst properties
on the time-variable physical conditions in the ignition area. For
example, the energetics and timescales of bursts depend on the
composition of the accreting material (see, e.g., Cumming \& Bildsten
2001), the local rate of accretion, the temperature of the
neutron-star core (e.g., Fushiki \& Lamb 1987), the presence of ashes
from previous bursts, etc. Recently, there has been significant
progress towards developing models of thermonuclear flashes that relax
many of the limiting assumptions of earlier calculations and
incorporate detailed nuclear networks (Schatz et al.\ 2000), the
effects of multi-dimensional propagation of burning fronts (Zingale et
al.\ 2001; see also Fig.~\ref{fig:burn}), and the stellar rotation
(Spitkovsky et al.\ 2002).

The long-term monitoring capabilities of {\em BeppoSAX\/} and {\em
RXTE} brought the discovery of a new type of long ($\sim $hr) bursts
with even longer ($\sim$yr) recurrence times, the so-called
superbursts (Cornelisse et al.\ 2000; Strohmayer \& Bildsten, \S3). These are believed to be caused by unstable carbon burning in
layers that are deeper than those responsible for the normal Type~I
bursts, thereby accounting for their longer durations and recurrence
timescales (Strohmayer \& Brown 2002; Cumming \& Bildsten 2002).

During many of the normal Type~I X-ray bursts and in one superburst,
highly coherent oscillations of the observed X-ray fluxes are often
detected (Strohmayer et al.\ 1996; Strohmayer and Bildsten, this
volume). The frequencies of the oscillations drift by a few percent
during the bursts, reaching values that are constant, to within one
part in $10^4$, between bursts from the same source (Muno et al.\
2002a). In bursts from two ultracompact millisecond pulsars, in which
the spin frequencies of the stars are known, the asymptotic values of
the burst oscillation frequencies are nearly \mk equal to the spin
frequencies of the stars (see, e.g., Chakrabarty et al.\ 2003 and
\S2\mk).
% there are only two cases of this \mk

These two properties have led to an interpretation of burst
oscillations in which the thermonuclear burning on the neutron star
surface is non-uniform and produces a modulation of the X-ray flux at
the stellar spin frequency (Strohmayer et al.\ 1996). The origin of
the frequency drift during the rising phase of the bursts, however, is
still unresolved. A number of ideas are currently being explored,
which involve the decoupling and slowing down of the surface layers
from the rest of the neutron star during the rise (Cumming et al.\
2002), the drift of the burning front with respect to a fiducial
azimuth on the stellar surface caused by rotation (Spitkovsky et al.\
2001), and the excitation of non-radial modes in the burning layers
(Heyl 2004).

Despite the lack of a physical model of burst oscillations, their
properties have already been used in obtaining stringent constraints
on the masses and radii of neutron stars, as well as on the degree of
non-uniformity of the thermonuclear burning. For example, during burst
rise, the large ($\sim 70\%$) observed amplitudes of the oscillations
require that the neutron stars are not too compact for gravitational
self-lensing to wash out the pulsations (Nath et al.\ 2002). Moreover,
the lack of detectable harmonics in burst oscillations constrains \mk the
emission areas and orientations with respect to the rotation axes of
the stars (Muno et al.\ 2002b).

\subsection{A Census of Non-Pulsing Neutron Stars and Black Holes}

As it has become evident from the discussion in the previous section,
non-pulsing neutron stars and black holes are found in a variety of
binary systems and configurations. The number of known sources has
steadily increased, from 33 in 1983, to 119 in 1995, to 150 in 2000
(Liu et al.\ 2001). This increase is largely due to the discovery of
low-luminosity sources with detectors of increasing sensitivity, but
also due to the discovery of a large number of transient sources.

\begin{figure*}
 \centerline{ \psfig{file=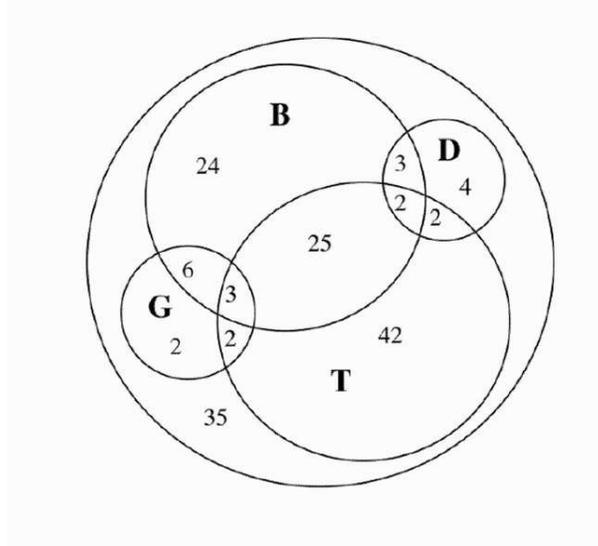,angle=0,width=8truecm}} 
\caption{A Venn diagram of the known non-pulsing neutron stars and black
holes in binary systems (using the catalog of Liu et al.\ 2001). The initials
correspond to B: bursters; G: globular-cluster sources; T: transients;
D: dippers. The areas of the circles correspond approximately to the
relative number of sources in each category.}  \label{fig:venn}
\end{figure*}

Out of 150 sources, 63 exhibit Type~I X-ray bursts and can be,
therefore, identified as neutron stars. On the other hand, 17
sources have dynamically measured masses in excess of $3.2 M_\odot$
and can be safely identified as black holes.  Half of the known
sources (76 out of 150) are transients. Given their relatively low
duty cycles, which is of order 10\% for neutron-star sources and much
lower for black holes, the total number of transient sources in the
galaxy must be significantly larger (King \& Kolb 1996).

Finally, about 10\% of the objects in each category (5/63 bursters,
4/76 transients, 11/150 total) are dippers. This is consistent with
the geometric interpretation of the dips and with an opening angle of
the accretion stream (as measured from the central objects) of about
10 degrees.
% how estimated? mk

\subsection{The Nature of the Central Object in Non-Pulsing X-ray Binaries}
\label{sect:nature}

The presence of persistent pulsations or of Type~I bursts in the X-ray
lightcurve of an accreting compact object provide the only unequivocal
proof that the central object is a compact {\em star\/} and not a
black hole. However, the absence of pulsations or bursts does not
provide proof that the compact object does not have a surface (see,
however, Narayan \& Heyl 2002). Indeed, if the magnetic field of the
neutron star is dynamically unimportant ($\ll 10^9$~G), it will not
appear as an X-ray pulsar.  Moreover, Type~I X-ray bursts are expected
to occur only for a particular range of accretion rates, surface
gravity accelerations, compositions of the accreting material, and
core temperatures (see, e.g., Fushiki \& Lamb 1987).  Finally, the
presence of pulsations or bursts indicates simply the existence of a
stellar surface and does not necessarily prove that the object is a
neutron star, as opposed to another type of compact star with
yet-to-be-discovered properties. It is in fact extremely difficult to
provide a conclusive proof that a non-pulsing object is a black hole
or even a neutron star. All current arguments are either empirical or
simply attempted proofs by elimination, leaving open the possibility
that a viable alternative was simply not considered (see also
Abramowicz et al.\ 2002).

Since the discovery of X-ray binaries, a number of empirical tests
have been put forward for distinguishing non-pulsing neutron stars
from black holes based on their X-ray properties. Recent examples of
suggested evidence for black holes are the presence of a hard X-ray
spectrum at high luminosities (Barret et al. 1996), of a particular
type of X-ray spectrum (Done \& Gierlinski 2003), or of significant
variability power at high Fourier frequencies (Sunyaev \& Revnivtsev
2000). Albeit useful as indicators, these empirical test cannot
provide conclusive arguments for the nature of the central
objects. Neutron stars and black holes share the general
characteristics of a very complex phenomenology of spectral and timing
properties and their differences are only in the details (see, e.g.,
van der Klis 1994 for a discussion).

The measurement of a large gravitational mass for the compact object
is currently considered to be the strongest evidence for its
identification with a black hole. The reason is that, under three
simple assumptions, an optimal upper bound on the mass of any neutron
star of $\simeq 3.2 M_\odot$ can be derived (Rhodes \& Ruffini
1983). The three \mk assumptions are: (i) the star is non-rotating; (ii)
the equation of state at densities below some fiducial value
(typically close to the nuclear saturation density) is known; (iii)
the speed of sound at larger densities is smaller than the speed of
light (the so-called ``causality'' condition). 

Including the effects of rotation introduces only small corrections
and affects the limiting mass by $<20\%$ (Friedman \& Ipser 1987).
The next two assumptions, however, are significantly more
constraining. The speed of sound, being a phase velocity, is not
bounded by relativity to be less than the speed of light. In fact, the
actual condition used is a causality requirement only for cold,
non-dispersive material. However, neutron-star matter can be both
dispersive and of non-zero temperature. Relaxing this condition and
allowing for rotation leads to bounds on the neutron star mass as
large as $\sim 14 M_\odot$ (Sabbadini \& Hartle 1977). Finally, the
compact stars under consideration may form a distinct family of
objects, which is not a continuation of the normal white-dwarf to
neutron-star sequence of equilibrium configurations 
% is there one? \mk
towards higher
central densities. If this is the case, then the second of the above
assumptions is irrelevant.

Families of compact objects that are not bound by gravity, such as
strange star (Witten 1984; Alcock et al.\ 1986) and Q stars (Bahcall
et al.\ 1990), have been recently constructed as potential
alternatives to neutron stars and black holes. Strange
stars can be made practically indistinguishable from neutron \mk stars
with respect to masses, radii, and maximum spin frequencies (see,
however, Glendenning 1997). On the other hand, Q stars can be
constructed to have masses as large as the most massive stellar-mass
\mk black-hole candidates,
even though this would require extreme changes in our understanding of
the properties of matter at densities as low as one tenth of nuclear
saturation (Miller et al.\ 1998). 
% so, is this not contradicted by ion collision experiments then? \mk
Finally, if gravity is not described
by general relativity in the strong-field regime, then the limiting
mass of a neutron star may not be $\sim 3.2 M_\odot$. Metric theories
of gravity that are consistent with all solar system tests 
% so? how about pulsars\mk
but deviate
from general relativity in \mk the strong-field regime allow for neutron
stars with significantly larger mass (see, e.g., DeDeo \& Psaltis
2003).

It is important to note here that the existence of black holes is a
strong-field prediction of a theory (i.e., general relativity) that
has been tested to high accuracy, at least in the weak-field limit
(Will 2000). On the other hand, all the other alternatives discussed
above are the results of theoretical assumptions that have not been
tested (and mostly could not have been tested) with current
experiments. Such alternatives provide physically consistent
counter-examples \mk to the identification of a compact object as a black
hole.  However, they will remain simply as thought experiments until
experimental evidence shows that our theories of gravity and matter
fail to describe extreme physical conditions.

\begin{figure*}
 \centerline{ \psfig{file=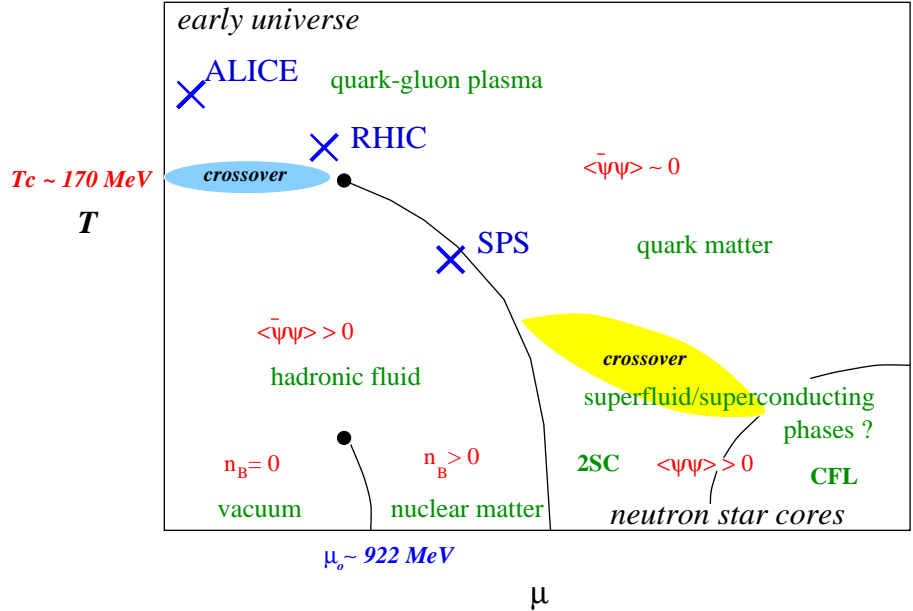,angle=0,width=12truecm}} 
\caption{The proposed phase diagram of QCD, showing the position on it of
matter in the early universe, in modern heavy-ion colliders (SPS,
RHIC, and ALICE), and in the cores of neutron stars (for details see
Hands 2001).}
\label{fig:qcd}
\end{figure*}

\section{Accretion-powered X-ray sources in the 21st century}

\begin{figure*}
 \centerline{ \psfig{file=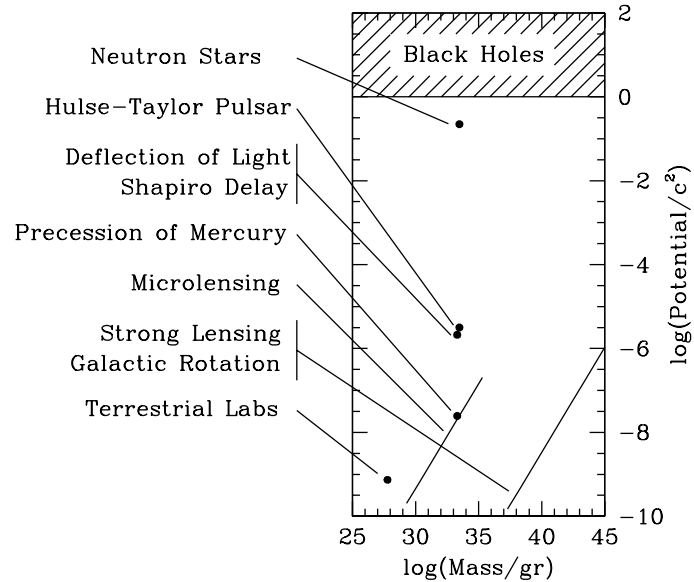,angle=0,width=10truecm}} 
\caption{The potential of the gravitational fields probed by different
astrophysical observations and tests of General Relativity (after
Psaltis 2004).}
\label{fig:gr_tests}
\end{figure*}

Accreting neutron stars and black holes in the galaxy offer the unique
opportunity of understanding the properties of matter,
electromagnetic, and gravitational fields beyond the conditions found
in current terrestrial experiments and other cosmic settings. Indeed,
matter in the cores of neutron stars occupies a place in the proposed
QCD phase diagram that is distinct from the regions occupied by matter
in the early universe and in modern heavy-ion colliders
(Fig.~\ref{fig:qcd}; Hands 2001). At the same time, the gravitational
fields probed by the accretion flows just outside the event-horizons
of black holes and the surfaces of neutron stars, are many orders of
magnitude stronger than those probed by other tests of general
relativity (Fig.~\ref{fig:gr_tests}; Psaltis 2004).  X-ray astronomy
and the discovery of accretion-powered neutron stars and black holes
provide probes with which tests of basic physics theories can be
performed in a way that is complementary to other experiments and
cosmological studies.

The last four decades have been the period of discovery, in which the
astrophysical properties of accreting compact objects were
investigated.  For his contribution to this effort, Riccardo Giaconni
was awarded the 2002 Nobel price in physics.  In the near future, the
observations of neutron stars and black holes with detectors \mk with large
surface areas, high spectral resolution, and fast timing capabilities
will allow for precise measurements of the physical conditions in the
accretion flows, as they vary at the dynamical timescales near the
compact objects. Moreover, the increase in the computational power and
storage capabilities of supercomputers will allow for the development
of new tools for modeling radiation-magneto-hydrodynamic phenomena in
curved spacetimes. And, as it has always been the case in
compact-object astrophysics, this interplay between theory and
observations will offer us a more complete picture of our universe.

{\bf Acknowledgements:} It is my pleasure to thank a number of people
that have helped me in understanding the concepts described in this chapter;
I thank especially T.\ Belloni, D.\ Chakrabarty, S.\ DeDeo,
D.\ Galloway, E.\ Kuulkers, F.\ Lamb, C.\ Miller, M.\ Muno,
R.\ Narayan, F.\ \"Ozel, and M.\ van der Klis. I am also
grateful to Deepto Chakrabarty for help in planning, writing, and
proofreading this chapter, to Duncan Galloway for producing several of
the figures, as well as to Martin Pessah and Erik Kuulkers for
carefully reading the manuscript.

\end{document}